\documentclass[twocolumn,showpacs,aps,prxquantum,letterpaper,superscriptaddress,nofootinbib,raggedbottom,nobalancelastpage]{revtex4-2}
\usepackage{graphicx,bm}
\usepackage{amsmath}
\usepackage{amssymb}
\usepackage{xcolor}
\usepackage{color}
\begin{document}
\title{Quantum gates utilizing dark and bright states in open dissipative cavity QED}
\author{Mikhail Tokman}
\affiliation{Department of Electrical and Electronic Engineering and Schlesinger Knowledge Center for Compact Accelerators
and Radiation Sources, Ariel University, 40700 Ariel, Israel}
\author{Jitendra K. Verma}
%\email{j.vermakumar@gmail.com}
\affiliation{Department of Physics and Astronomy, Texas A\&M University, College Station, TX, 77843 USA}
%\author{Jacob Bohreer}
%\affiliation{Department of Physics and Astronomy, Texas A\&M University, College Station, TX, 77843 USA}
\author{Alexey Belyanin}%
\email{belyanin@tamu.edu}
\affiliation{Department of Physics and Astronomy, Texas A\&M University, College Station, TX, 77843 USA}
\date{\today}% It is always \today, today, %  but any date may be explicitly specified
\begin{abstract}
{
We present a general formalism and specific implementation of quantum gates based on interaction of single photons with open dissipative nanocavities containing ensembles of quantum emitters. Rich dynamics of entangled bright and dark states of quantum emitters coupled to a nanocavity field gives rise to efficient manipulation of the quantum state of an incident photon. In its simplest implementation, an initial preparation of the state of quantum emitters by a classical optical pulse controls the polarization state of the reflected photon.   
}
\end{abstract}
%\pacs{03.65.Ud, 03.67.Mn, 42.50.Dv}
\maketitle

\section{Introduction}

Cavity QED utilizing strong coupling and entanglement between quantum emitters (QEs) and a quantized EM cavity mode has been recognized a long time ago as an attractive platform for quantum information applications \cite{haroche2006}.  A more recent development of solid-state dielectric and plasmonic nanocavities integrated with quantum dots or defect-based QEs and coupled to nanophotonic circuits capable of operating in the near-IR telecommunications range promises chip-compatible and scalable quantum information systems \cite{thorma2015, lodahl2015, lukin2016, dovzhenko2018,bitton2019,wang2020}.  However, scaling down the cavity size, nanophotonic integration, and solid-state environment for QEs lead to increased losses for a cavity field and faster decoherence for QE-based qubits. Furthermore, an access to QEs buried in the semiconductor or dielectric crystal is limited, complicating the use of their quantum states in quantum gates. 

An interesting possibility which avoids many of the above detrimental effects is to treat the whole cavity loaded with QEs as a logical qubit which interacts with an external incident photon serving as a flying qubit. The degrees of freedom of a photon (e.g., its polarization) are affected as it scatters or reflects off of a cavity, or interacts with a cavity while propagating in an evanescently coupled waveguide. This interaction can be controlled by changing the state of the QEs in a cavity with external classical electromagnetic fields. This scenario of implementing gate operations has been studied a number of times for a single QE in a cavity \cite{hughes2004, jung2009, hacker2016, reiserer2013, waks2013,waks2018}, including experimental realization in an integrated solid-state setting \cite{waks2013,waks2018}, in which the quantum dot states in a photonic crystal nanocavity were split by a strong magnetic field to create a state detuned  from the cavity mode.  The reflection of a single photon from the cavity will then depend on whether the quantum dot is in the state detuned from or strongly coupled to the cavity mode. 

It is well known that many-body eigenstates in an ensemble of two or more QEs are split into bright and dark entangled states with respect to the coupling to a cavity field; see, e.g.,  \cite{evans2018,  koong2022, tokman2023, lei2023}. These states have different symmetry, but they may also have different energies due to, e.g., the dipole-dipole interaction which is inevitable in a nanocavity. This makes the control of the ``cavity qubit'' particularly easy without the need for any additional magnetic fields: a classical optical field can toggle the QEs between the ground (strongly coupled) state and the dark (decoupled) state, thus changing the polarization of the reflected photon. A variety of regimes is possible, depending on the optimization with respect to the system parameters.  

The above optimization and the general analysis of this problem are nontrivial due to the need to include dissipation and fluctuations for all degrees of freedom, as well as coupling of an external photon to the cavity field. The large number of parameters makes a numerical analysis, based, e.g., on the Lindblad master equation, not very useful and predictive. Recently we have developed a new formalism based on the stochastic Schroedinger-Langevin approach which allows one to obtain analytic solutions for the nonperturbative quantum dynamics of the open systems with many quantized degrees of freedom, including dynamic coupling to phonons or vibrational modes \cite{tokman2021,tokman2022}, time-dependent parameters \cite{chen2021}, highly nonuniform quantized nanocavity fields \cite{tokman2023}, a large spread of transition frequencies in the ensemble of strongly coupled quantum emitters \cite{tokman2023}, and the effects of external photon propagation and coupling \cite{tokman-prl2023}. Developing this formalism for the present problem, we obtain the analytic solution for the quantum state of the reflected photon depending on the state of QEs and relaxation and coupling parameters showing the possibility of a two-qubit gate operation. 

The paper is structured as follows. Sec.~\ref{section2} details the model of an external single-photon state interacting with an cavity containing a quantized cavity filed coupled to two QEs. Sec.~\ref{section3} solves the quantum dynamics of this system in the presence of dissipation and noise for both the cavity and QEs. In Sec.~\ref{section4} we describe the operation of a two-qubit gate utilizing the polarization state of the reflected photon as a signal. The discussion and conclusions are in Sec.~\ref{section5}.  The Appendix provides the derivation details.

%%%%%%%%%%%%%%%%%%%%%%%%%%%%%%%%%

\section{Coupling of external photons to an open dissipative cavity: the model} 
\label{section2}

We consider the simplest situation which still allows many-body dark and bright states: two 2-level QEs at positions $\mathbf{r}_{1}$ and $\mathbf{r}_{2}$ in an anisotropic cavity, see the sketch in Fig.~1. Generalization to $N$ QEs is straightforward  \cite{tokman2023}. The wave functions of QEs do not overlap, but dipole-dipole interaction is possible and even inevitable in a nanocavity. 

%%%%%%%%%%%%%%%%%%%%%%%%%%%%%%%%%

\begin{figure}[h!]
\centering
\includegraphics[width=8.5cm,height=4.5cm]{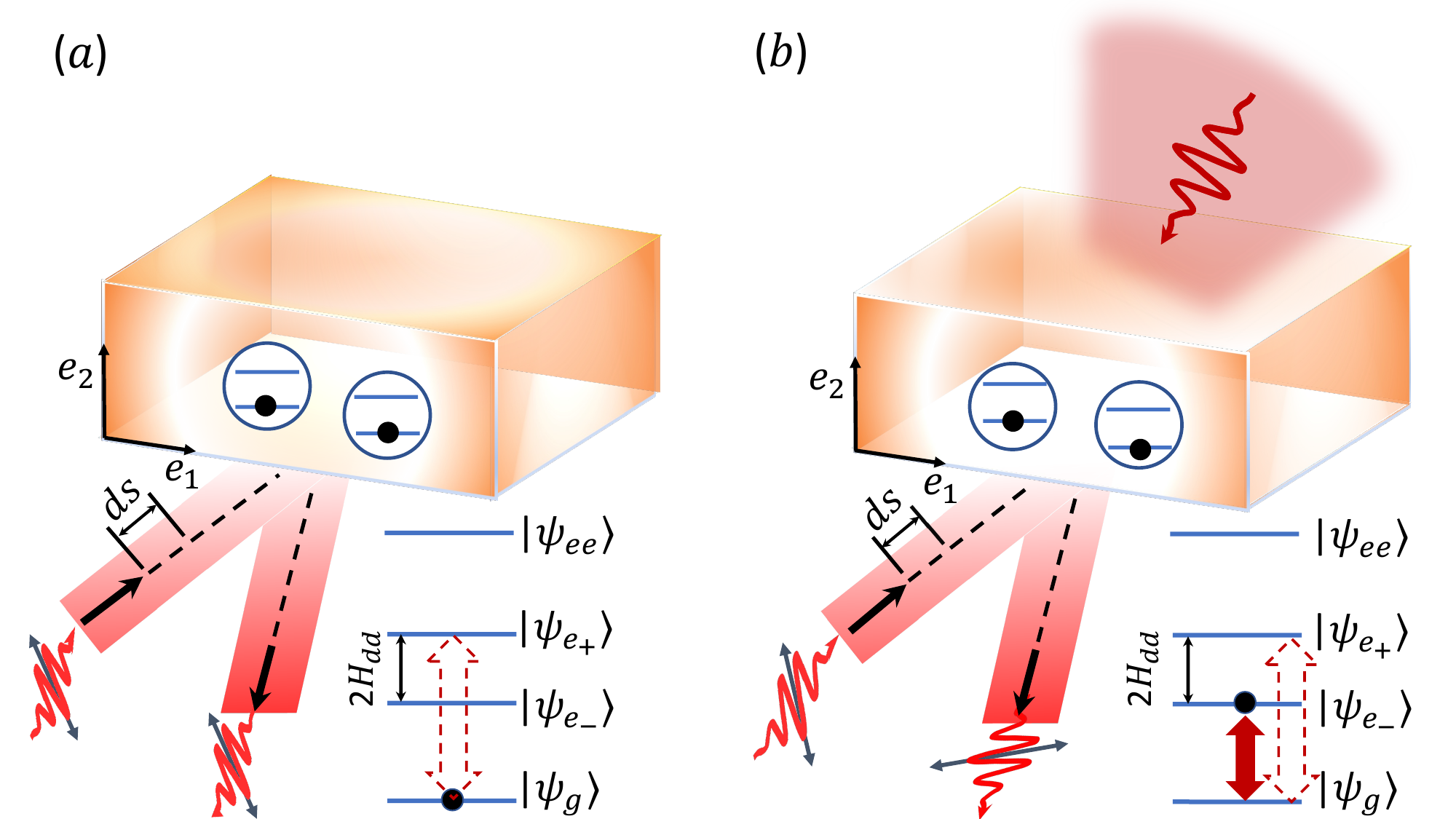}
\vspace{-0.1cm}
\caption{A schematic of an incident single photon state interacting with a cavity which contains two QEs. The energy levels of the eigenstates (\ref{eqn12}) of the QEs including their dipole-dipole interaction are indicated. As one implementation of the two-qubit gate, in (a) the QEs are in the ground state strongly coupled to the cavity mode (dashed double arrow), and the polarization state of a photon reflected through the cavity remains unchanged. In (b) the two-QE system is prepared by a classical field (filled arrow) in the dark state decoupled from the cavity mode, and the polarization of the reflected photon is rotated by $\frac{\pi}{2}$ with respect to incident photon.}
\label{fig1}
\end{figure}

%%%%%%%%%%%%%%%%%%%%%%%%%%%%%%%

The states of individual QEs are $|0_{j}\rangle$ and
$|1_{j}\rangle$ and their transition energies are $W_j$, where $j=1,2$. The dipole moment operators are ${\hat{\mathbf{d}}}_{j}=\mathbf{d}_{10}^{(j)}|1_{j}\rangle\langle 0_{j}| + \mathbf{d}_{01}^{(j)}|0_{j}\rangle\langle1_{j}|$, where $\mathbf{d}_{10}^{(j)}=\langle 1_{j}|{\hat{\mathbf{d}}}_{j}|0_{j}\rangle$,
$\mathbf{d}_{01}^{(j)}=\langle 0_{j}|{\hat{\mathbf{d}}}_{j}|1_{j}\rangle$. Introducing lowering and raising  operators $
{\hat{\sigma}}_{j} = |0_{j}\rangle\langle 1_{j}|$, $
{\hat{\sigma}}_{j}^{\dagger} = |1_{j}\rangle\langle 0_{j}|$,  
the dipole operators can be written as 
$ {\hat{\mathbf{d}}}_{j}=\mathbf{d}_{j}{\hat{\sigma}}_{j}^{\dagger} + \mathbf{d}_{j}^{\mathbf{\ast}}{\hat{\sigma}}_{j}$,
where $\mathbf{d}_{j} \equiv \mathbf{d}_{10}^{(j)}  $.
Introducing the operator of dipole-dipole interaction 
$\hat{H}_{dd} = \alpha_{dd}{\hat{\mathbf{d}}}_{1}{\hat{\mathbf{d}}}_{2}$, 
the Hamiltonian of QEs is 
\begin{equation}
{\hat{H}}_{e} = \sum_{j = 1,2} W_j {\hat{\sigma}}_{j}^{\dagger}{\hat{\sigma}}_{j} +\hbar\Omega_{dd}\left({\hat{\sigma}}_{1}^{\dagger}{\hat{\sigma}}_{2} + {\hat{\sigma}}_{1}{\hat{\sigma}}_{2}^{\dagger}\right),
\label{eqn4}
\end{equation}
where $\Omega_{dd} = \frac{\alpha_{dd}\mathbf{d}_{1}\mathbf{d}_{2}}{\hbar}$, which can be treated as real without loss of generality. Here we already assumed the rotating wave approximation (RWA) and put $W_1 = W_2$ in the interaction Hamiltonian to simplify the terms. Next, we introduce a 2-mode electric optical field of a cavity, which is a sum of symmetric (s) and antisymmetric (a) modes,
\begin{equation}
\hat{\mathbf{E}}=\sum_{i = a,s}\left(\mathbf{E}_{i}\left(\mathbf{r}\right){\hat{c}}_{i} + \mathbf{E}_{i}^{\mathbf{\ast}}\left(\mathbf{r}\right){\hat{c}}_{i}^{\dagger}\right),
\label{eqn5}
\end{equation}
where the fields $\mathbf{E}_{s,a}\left(\mathbf{r}\right)$ satisfy the normalization condition \cite{tokman2-2022} 
\begin{equation}
\int_{V}{\mathbf{E}_{i}}^{\mathbf{\ast}}\left(\mathbf{r}\right)\frac{\partial\big[\omega^{2}\overset{\leftrightarrow}{\varepsilon}(\omega,\mathbf{r})\big]}{\omega\partial\omega}\mathbf{E}_{i}\left(\mathbf{r}\right)d^{3}r= 4\pi\hbar\omega_{i}
\label{eqn6}
\end{equation}
and
\begin{equation}
\mathbf{E}_{s}\left(\mathbf{r}_{1}\right)=\mathbf{E}_{s}\left(\mathbf{r}_{2}\right),\
\mathbf{E}_{a}\left(\mathbf{r}_{1}\right)=-\mathbf{E}_{a}\left(\mathbf{r}_{2}\right). 
\label{eqn7}
\end{equation}
The spatial structure $\mathbf{E}_{s,a}\left(\mathbf{r}\right)$
and frequencies $\omega_{a,s}$ are derived from the classical
boundary-value problem, V is the quantization volume, e.g., the volume of a cavity,
$\overset{\leftrightarrow}{\varepsilon}\left(\omega,\mathbf{r}\right)$
is the dielectric tensor of a nonuniform dispersive medium. For
normalization (\ref{eqn6}) the field Hamiltonian has a standard form
\begin{equation}
\hat{H}_{f} = \hbar\sum_{i = a,s}{\omega_{i}\left({\hat{c}}_{i}^{\dagger}{\hat{c}}_{i} + \frac{1}{2}\right)},
\label{eqn8}
\end{equation}
where the creation and annihilation operators satisfy standard commutation relations
$\lbrack{\hat{c}}_{i},{\hat{c}}_{i^{\prime}}^{\dagger}\rbrack = \delta_{ii^{\prime}}$.
The Hamiltonian of the electric-dipole interaction with QEs is
\begin{equation}
{\hat{H}}_{ef}= -\int_{V}\hat{\mathbf{d}}\hat{\mathbf{E}}\, d^{3}r,
\label{eqn9}
\end{equation}
where $\hat{\mathbf{d}}=\sum_{1,2}{\hat{\mathbf{d}}}_{j}\delta\left(\mathbf{r}-\mathbf{r}_{j}\right)$.
Next we consider identical QEs with $W_{1} = W_{2} = W$ and
$\mathbf{d}_{1}=\mathbf{d}_{2}=\mathbf{d}$. Using
the QE Hamiltonian from Eq.~(\ref{eqn4}), from
$(\hat{H}_{e}-E)|\Psi\rangle=0$
we obtain the new eigenstates, 
\begin{eqnarray}
\left|\Psi_{g}\right\rangle = \left|0_{1}\right\rangle\left|0_{2}\right\rangle,\
\left|\Psi_{ee}\right\rangle = \left|1_{1}\right\rangle\left|1_{2}\right\rangle,\nonumber\\
\left|\Psi_{e_{\pm}}\right\rangle = \frac{\left|1_{1}\right\rangle\left|0_{2}\right\rangle\pm \left|0_{1}\right\rangle\left|1_{2}\right\rangle}{\sqrt{2}}
\label{eqn12}
\end{eqnarray}
and eigenenergies $E = 0$, $2W$, and  $W_{\pm} = W\pm\hbar\Omega_{dd}$, respectively, as sketched in Fig.~1. 
%\begin{equation}
%\begin{pmatrix}
%\left(2W - E\right) & 0& 0 &0\\
% 0 &\left(W_{+} - E\right) &0 & 0 \\
% 0 & 0 & \left(W_{-} - E \right) &0 \\
% 0 & 0 & 0 & -E
%\end{pmatrix}\times
%\begin{pmatrix}
%\Psi_{ee} \\
% \Psi_{e_{+}} \\
% \Psi_{e_{-}} \\
% \Psi_{g} \\
% \end{pmatrix} = 0
% \label{eqn10}
% \end{equation}
This results in
\begin{equation}
\hat{H}_{e} = 2W{\hat{\sigma}}_{ee}^{\dagger}{\hat{\sigma}}_{ee} + \sum_{\pm}W_{\pm}{\hat{\sigma}}_{e_{\pm}}^{\dagger}{\hat{\sigma}}_{e_{\pm}},
\label{eqn13}
\end{equation}
where we introduced the projection operators in the new basis,
\begin{eqnarray}
{\hat{\sigma}}_{ee} = \left|\Psi_{g}\right\rangle\left\langle\Psi_{ee}\right|,\
{\hat{\sigma}}_{ee}^{\dagger} = \left|\Psi_{ee}\right\rangle\left\langle\Psi_{g}\right|,\nonumber\\
{\hat{\sigma}}_{e_{\pm}} = \left|\Psi_{g}\right\rangle\left\langle\Psi_{e_{\pm}}\right|,\
{\hat{\sigma}}_{e_{\pm}}^{\dagger} = \left|\Psi_{e_{\pm}}\right\rangle\left\langle\Psi_{g}\right|. 
\label{eqn14}
\end{eqnarray}
The interaction Hamiltonian (\ref{eqn9}) expressed through the projection operators (\ref{eqn14}) becomes
\begin{eqnarray}
{\hat{H}}_{ef} &=& -\hbar\left\lbrack\Omega_{s}{\hat{c}}_{s}\left({\hat{\sigma}}_{e_{+}}^{\dagger} + {\hat{\Sigma}}_{e_{+}}^{\dagger}\right) + \Omega_{a}{\hat{c}}_{a}\left({\hat{\sigma}}_{e_{-}}^{\dagger} + {\hat{\Sigma}}_{e_{-}}^{\dagger}\right)\right\rbrack \nonumber \\ &+& {\rm H.c.},
\label{eqn15}
\end{eqnarray}
where ${\hat{\Sigma}}_{e_{\pm}} = {\hat{\sigma}}_{e_{\pm}}^{\dagger}{\hat{\sigma}}_{ee} = |\Psi_{e_{\pm}}\rangle\langle\Psi_{ee}|$, $\Omega_{s,a} = \sqrt{2}\frac{\mathbf{d}\mathbf{E}_{s,a}\left(\mathbf{r}_{1}\right)}{\hbar}$
is the cavity mode Rabi frequency and the factor $\sqrt{2}$ is due to two
QEs. The total Hamiltonian is 
${\hat{H} = \hat{H}}_{e} + {\hat{H}}_{f} + {\hat{H}}_{ef}$, 
%\begin{eqnarray}
%\hat{H} = \hbar\sum_{i=a,s}{\omega_{i}\left({\hat{c}}_{i}^{\dagger}{\hat{c}}_{i} + \frac{1}{2}\right)}+2W{\hat{\sigma}}_{ee}^{\dagger}{\hat{\sigma}}_{ee} + \sum_{\pm}{W_{\pm}{\hat{\sigma}}_{e_{\pm}}^{\dagger}{\hat{\sigma}}_{e_{\pm}}}\nonumber\\
%-\hbar\left\{\left\lbrack\Omega_{s}{\hat{c}}_{s}\left({\hat{\sigma}}_{e_{+}}^{\dagger} + {\hat{\Sigma}}_{e_{+}}^{\dagger}\right)+\Omega_{a}{\hat{c}}_{a}\left({\hat{\sigma}}_{e_{-}}^{\dagger}+ {\hat{\Sigma}}_{e_{-}}^{\dagger}\right)\right\rbrack + h.c.\right\}\nonumber\\
%\label{eqn16}
%\end{eqnarray}
%\begin{equation}
%\hat{\Sigma}_{e_\pm} = \hat{\sigma}_{e_\pm}^{\dagger}{\hat{\sigma}}_{ee} = \left. \ |\Psi_{e_\pm} \right\rangle\left\langle \Psi_{ee}|\right.\ 
%\end{equation}
which becomes in the interaction picture, 
\begin{eqnarray}
\hat{H} = - \hbar\Omega_{s}\hat{c}_{s}e^{-i\omega_{s}t}\left({\hat{\sigma}}_{e_{+}}^{\dagger}e^{i\frac{W_{+}}{\hbar}t} + {\hat{\Sigma}}_{e_{+}}^{\dagger}e^{i\frac{W_{-}}{\hbar}t}\right)\nonumber\\ -\hbar\Omega_{a}{\hat{c}}_{a}e^{-i\omega_{a}t}\left( {\hat{\sigma}}_{e_{-}}^{\dagger}e^{i\frac{W_{-}}{\hbar}t} + {\hat{\Sigma}}_{e_{-}}^{\dagger}e^{i\frac{W_{+}}{\hbar}t}\right) + \rm H.c.
\label{eqn17}
\end{eqnarray}
Here we took into account that $2W - W_{\pm} = W_{\mp}$. Note that if we want to use the Hamiltonian (\ref{eqn17}) to calculate the observables, we need to make the following substitutions for the operators:
${\hat{c}}_{s,a}\Longrightarrow{\hat{c}}_{s,a}e^{-i\omega_{s,a}t},
{\hat{\sigma}}_{e_{\pm}}\Longrightarrow{\hat{\sigma}}_{e_{\pm}}e^{-i\frac{W_{\pm}}{\hbar}t},
{\hat{\sigma}}_{ee}\Longrightarrow{\hat{\sigma}}_{ee}e^{-i\frac{2W}{\hbar}t},
{\hat{\Sigma}}_{e_{\pm}}\Longrightarrow{\hat{\Sigma}}_{e_{\pm}}e^{-i\frac{W_{\mp}}{\hbar}t}$.

%%%%%%%%%%%%%%%%%%%%%%%%%%%%%%%%%%%%%%%%%%%

\subsection{Control of QE states with a classical EM field} 

It is obvious that a strong classical EM field pulse of the right symmetry, frequency, and duration, incident at an open cavity, can bring the QEs into a desired symmetric or antisymmetric state. We add this subsection below just for completeness and to mention the restrictions on the pulse parameters. 

Consider a classical field which has both symmetric and antisymmetric modes present,
\begin{equation}
\mathbf{E} = \sum_{i=a,s}{\mathcal{E}_{i}\left(\mathbf{r}\right)e^{-i\omega_{i}t}} + c.c.,
\label{eqn18}
\end{equation}
where $\mathcal{E}_{i}\left(\mathbf{r}\right)$ are complex
field amplitudes. The coupling to this field adds the terms to the Hamiltonian (\ref{eqn17})  in which $\Omega_{s}{\hat{c}}_{s}$ and
$\Omega_{a}{\hat{c}}_{a}$ are replaced by
classical Rabi frequencies, $\Omega_{s,a}{\hat{c}}_{s,a}\Longrightarrow\Omega_{s,a}^{(cl)} = \sqrt{2}\frac{\mathbf{d}\mathcal{E}_{s,a}\left(\mathbf{r}_{1}\right)}\hbar$.
%\begin{eqnarray}
%\hat{H} = - \hbar\Omega_{s}^{(cl)}e^{-i\omega_{s}t}\left({\hat{\sigma}}_{e_{+}}^{\dagger}e^{i\frac{W_{+}}{\hbar}t} + {\hat{\Sigma}}_{e_{+}}^{\dagger}e^{i\frac{W_{-}}{\hbar}t}\right)\nonumber\\ - \hbar\Omega_{a}^{(cl)}e^{-i\omega_{a}t}\left({\hat{\sigma}}_{e_{-}}^{\dagger}e^{i\frac{W_{-}}{\hbar}t} + {\hat{\Sigma}}_{e_{-}}^{\dagger}e^{i\frac{W_{+}}{\hbar}t}\right) + h.c.
%\label{eqn19}
%\end{eqnarray}
If a resonant optical field with only type of symmetry is present, one can excite only one kind of the QE state. For definiteness, let it be the field of an 
\emph{a}-type at frequency $\omega_{a} = \frac{W_{-}}\hbar$, as sketched with a solid arrow in Fig.~\ref{fig1}. 
The same analysis could be repeated for an $s$-type field at
frequency $\omega_{s} = \frac{W_{+}}\hbar$.
Neglecting the nonresonant terms, the Hamiltonian becomes
\begin{equation}
\hat{H}\approx - \hbar\Omega_{a}^{(cl)}{\hat{\sigma}}_{e_{-}}^{\dagger} + h.c.
\label{eqn20}
\end{equation}
It corresponds to the state vector
\begin{equation}
|\Psi(t)\rangle = C_{g}(t)|\Psi_{g}\rangle + C_{e_{-}}(t)|\Psi_{e_{-}}\rangle,
\label{eqn21}
\end{equation}
where $C_{g}$ are $C_{e_{-}}$ are corresponding probability
amplitudes. In this case the Schrodinger equation gives
\begin{equation}
\begin{pmatrix}
\frac{d}{dt} & -i\Omega_{a}^{(cl)} \\
-i{\Omega_{a}^{(cl)}}^{\ast} & \frac{d}{dt} \\
\end{pmatrix}
\begin{pmatrix}
C_{e_{-}} \\
C_{g}\\
\end{pmatrix} = 0
\end{equation}
So that for the initial conditions $C_{g}(t=0)=1$, $C_{e_{-}}=0$ the solution is
\begin{equation}
\begin{pmatrix}
C_{e_{-}}\\\\
C_{g} \\
\end{pmatrix} = \begin{pmatrix}i\sin\left(\Omega_{a}^{(cl)}t\right)\\
\cos\left(\Omega_{a}^{(cl)}t\right)
\end{pmatrix}.
\label{eqn22}
\end{equation}
Therefore, a square pulse of duration $\Omega_{a}^{\left(cl\right)}t = \frac{\pi}{2}$ brings the
system from the ground state to state $|\Psi_{e_{-}}\rangle$. The latter state will be
completely decoupled (dark) for the symmetric field. In the same way one can transfer the system to state $|\Psi_{e_{+}}\rangle$, by acting with a pulse of an symmetric field at frequency $\omega_{s} = \frac{W_{+}}\hbar$. Of course we invoked constant field amplitudes or square pulses just for analytic illustration. For realistic pulses of, e.g., Gaussian shape one can solve the same equations numerically with the same qualitative outcome. We also assumed pulse duration much shorter than the relaxation times. Furthermore, the field amplitude should be low enough, 
$|\Omega_{s,a}^{\left(cl\right)}| \ll {2\Omega}_{dd}$, to prevent excitation of both symmetric and antisymmetric states $|\Psi_{e_{\pm}}\rangle$ at the same time, or even the highest state  $|\Psi_{ee}\rangle$. 

%%%%%%%%%%%%%%%%%%%%%%%%%%%%%%%%%%%%%%

\subsection{Interaction of a cavity with an incident quantized field} 

As the last ingredient of the two-qubit gate, consider an incident 
quantum field which can couple to the cavity mode if the former has a nonzero projection onto the cavity polarization axis. We assume for definiteness that the quantized cavity field is single mode, 
\begin{equation}
\hat{\mathbf{E}}_{c}=\mathbf{E}_{c}\left(\mathbf{r}\right)\hat{c}_{c} + \mathbf{E}_{c}^{\mathbf{\ast}}\left(\mathbf{r}\right){\hat{c}}_{c}^{\dagger},
\label{eqn23}
\end{equation}
and has $s$-symmetry enabling strong coupling to the transition between the ground state $|\Psi_g\rangle$ and the symmetric state $|\Psi_{e_{+}}\rangle$. The normalization field $\mathbf{E}_{c}\left(\mathbf{r}\right)$
satisfies condition (\ref{eqn6}), where \emph{V} is the cavity volume.

%If the classical field has a large enough amplitude,
%$|\Omega_{s,a}^{\left(cl\right)}|\geq{2\Omega}_{dd}$,
%it can excite not only the state corresponding to its symmetry, i.e.
%$|\Psi_{e_{+}}\rangle$ or
%$|\Psi_{e_{-}}\rangle$, but also the highest state
% $|\Psi_{ee}\rangle$ (see Fig.\ref{fig1}).
%However, if state $|\Psi_{g}\rangle$ is empty and only
%one of the states $|\Psi_{e_{\pm}}\rangle$ or
%$|\Psi_{ee}\rangle$ are excited, a weak enough field of the opposite
%symmetry and right frequency won't couple to QEs. The
%condition on the field amplitude is
%$|\Omega_{s,a}^{\left(cl\right)}|\ll{2\Omega}_{dd}$
%for a classical field and $\sqrt{n_{ph}}\left|\Omega_{s,a}\right|\ll{2\Omega}_{dd}$
%for a quantum field with photon number $n_{ph}$.

The spatial structure of the external quantum field outside the cavity can be described as a  ray
bundle incident on the cavity and reflected from it, as sketched in Fig.~\ref{fig1}.
The quantization volume for the outside field includes both incoming and outgoing segments of
this ray bundle. The quantized modes are defined by the periodic
boundary conditions at the input and output facets of the beam, and
ideal reflection of the illuminated side of the cavity. The ideal
reflection condition is just the method of defining the set of quantized
modes which can be used as a basis for expansion of any field. This does
not contradict the fact that the field can be coupled to the cavity,
absorbed etc.: these processes can be described as the coupling between
a given cavity mode and the quantized modes of an external field. This
approach is valid if the coupling causes a relatively small change in
the frequency of the cavity mode and the spatial structure of the field.

It is convenient to 
introduce the coordinate \emph{s}, where $ds$ is the line
element along the central axis of the beam in Fig.~\ref{fig1}. The incident field
corresponds to $s < 0$, and the reflected one is for $s > 0$.
Outside the small region where the incident and reflected fields
overlap, the operator of an external field can be written as
\begin{equation}
\hat{\mathbf{E}}_{out}\left(\mathbf{\xi}_{\mathbf{\bot}},s\right)=\sum_{k} \sum_{p = 1,2} \mathbf{e}_{p}E_{k}\left(\mathbf{\xi}_{\mathbf{\bot}},s\right){\hat{c}}_{kp}e^{iks}+h.c.,
\label{eqn24}
\end{equation}
where the set of wavenumbers includes only positive \emph{k},
$\mathbf{e}_{1,2}$ are the polarization vectors of the two normal
modes which lie in the plane orthogonal to the central ray; a set of
wavenumbers \emph{k} is determined by periodic boundary conditions:
$kL = 2{\pi}N$ where \emph{L} is the length of the ray bundle;
$\mathbf{\xi}_{\mathbf{\bot}}$ and
$|E_{k}\left(\mathbf{\xi}_{\mathbf{\bot}},s\right)|^{2}$
are the coordinates and the intensity distribution in the plane
orthogonal to the central ray; $\hat{c}_{k1,2}$ are
annihilation operators of the Fock states for these normal modes.
Alternatively, we could choose the quantization volume for an external
field in the shape of a cylinder with \textbf{\emph{z}} normal to the
illuminated surface and introduce two kinds of quantized modes:
$\mathbf{\propto}\cos{kz},\sin{kz}$ or
$\propto e^{ikz},e^{- ikz}$. This would give the same
results but make the calculations more cumbersome. Moreover, our present approach is also applicable to the integrated circuit geometry in which an external pulse propagates in the waveguide and is evanescently coupled to the cavity, as in \cite{tokman-prl2023}.

Assuming that the photons are incident from the space with lower refractive index than that of a cavity, we
 include the $\pi$-shift upon reflection, 
\begin{equation}
E_{k}\left(\mathbf{\xi}_{\mathbf{\bot}},s < 0 \right) = E_{k}\left(\mathbf{\xi}_{\mathbf{\bot}} \right),
E_{k}\left(\mathbf{\xi}_{\mathbf{\bot}},s > 0 \right) = E_{k}\left(\mathbf{\xi}_{\mathbf{\bot}} \right)e^{{i\pi}}.
\label{eqn25}
\end{equation}
Furthermore, we assume that the outside medium is isotropic and the
normal modes are degenerate with respect to polarization, i.e., the
field can be split into two orthogonal polarizations in an arbitrary
way. Then the normalization condition for
$E_{k}\left(\mathbf{\xi}_{\mathbf{\bot}} \right)$ becomes
\begin{equation}
\int_{S_{\bot}}{\left\{\frac{\partial\left\lbrack \omega^{2}\varepsilon\left(\omega,\mathbf{\xi}_{\mathbf{\bot}}\right) \right\rbrack}{\omega\partial\omega}\right\}_{\omega = \omega_{k}}\left| E_{k}\left( \mathbf{\xi}_{\mathbf{\bot}} \right) \right|^{2}d^{2}\xi} = \frac{4\pi}{L}\hbar\omega_{k},
\label{eqn26}
\end{equation}
where $S_{\bot}$ is the cross-sectional area of the ray bundle,
$\varepsilon\left( \omega,\mathbf{\xi}_{\mathbf{\bot}} \right)$ is the
dielectric function of a medium inside the ray bundle. The functions
$E_{k}\left( \mathbf{\xi}_{\mathbf{\bot}} \right)$ are the same for
both orthogonal polarizations. The latter assumption and the mutual
orthogonality of the polarization vectors are approximate for a beam of
finite width. Note that in the limit $kL\gg{2\pi}$, which
corresponds to the limit of a continuous spectrum, the observables
should not depend on the quantization length $L$. 

Next, we assume that the cavity is 
polarization-selective, i.e., only the mode with polarization $\mathbf{e}_{1}$ 
can excite the cavity field, whereas the mode $\mathbf{e}_{2}$ is reflected with reflectivity 1. As
a result, the Hamiltonian becomes 
\begin{eqnarray}
\hat{H} = &\sum_{k}\left\lbrack\hbar\omega_{k}\left(\hat{c}_{1k}^{\dagger}\hat{c}_{1k} + \frac{1}{2}\right) - \hbar\left({\mathfrak{L}_{k}\hat{c}}_{1k}{\hat{c}}_{c}^{\dagger} + h.c.\right)\right\rbrack\nonumber\\
 &+\hbar\omega_{c}\left(\hat{c}_{c}^{\dagger}\hat{c}_{c} + \frac{1}{2}\right)+ W_{e}{\hat{\sigma}}_{e}^{\dagger}{\hat{\sigma}}_{e} - \hbar\left(\Omega_{c}{\hat{c}}_{c}{\hat{\sigma}}_{e}^{\dagger}+h.c.\right)\nonumber\\
 &+\sum_{k}\hbar\omega_{k}\left(\hat{c}_{2k}^{\dagger}\hat{c}_{2k} + \frac{1}{2}\right).
 \label{eqn27}
\end{eqnarray}
%\begin{eqnarray}
%\hat{H} = \sum_{k}\hbar\omega_{k}\left(\hat{c}_{1k}^{\dagger}\hat{c}_{1k} + \frac{1}{2}\right) +\sum_{k}\hbar\omega_{k}\left(\hat{c}_{2k}^{\dagger}\hat{c}_{2k} + \frac{1}{2}\right)\nonumber\\ - \hbar\left(\Omega_{c}{\hat{c}}_{c}{\hat{\sigma}}_{e}^{\dagger}+h.c.\right)-\hbar\left({\mathfrak{L}_{k}\hat{c}}_{1k}{\hat{c}}_{c}^{\dagger} + h.c.\right)\nonumber\\
% +\hbar\omega_{c}\left(\hat{c}_{c}^{\dagger}\hat{c}_{c} + \frac{1}{2}\right)+ W_{e}{\hat{\sigma}}_{e}^{\dagger}{\hat{\sigma}}_{e}\nonumber\\.
% \label{eqn27}
%\end{eqnarray}
Here $W_{e} = W_{\pm}$ and
$\hat{\sigma}_{e} = \hat{\sigma}_{e_{\pm}}$ depending on the symmetry of the field;
$\Omega_{c} = \sqrt{2}\frac{\mathbf{d}\mathbf{E}_{c}\left(\mathbf{r}_{1}\right)}{\hbar}$
is the Rabi frequency and we assume that $|\Omega_{c}| < {2\Omega}_{dd}$;
$\hbar\mathfrak{L}_{k}$ is the coupling energy of the cavity mode
with the mode of incident field of polarization $\mathbf{e}_{1}$ and
wavenumber \emph{k}. For simplicity we will assume that the coupling strength
does not depend on $k$, $\mathfrak{L}_{k}\equiv\mathfrak{L}$, similarly to \cite{jung2009}.

%%%%%%%%%%%%%%%%%%%%%%%%%%%%%%%%%%%%%%%%%%%%%%%%%%%
\section{Quantum dynamics of an external photon-cavity interaction} 
\label{section3}

Here we solve the quantum dynamics of the system described by the Hamiltonian (\ref{eqn27}), taking into account relaxation and noise within the stochastic Schroedinger equation (SSE) approach. For a single-photon incident field, including only the resonant terms and the
ground state, we seek the solution for the state vector as 
%\begin{widetext}
%\begin{eqnarray}
%|\Psi\rangle = |0_{c}\rangle{|\Psi}_{g}\rangle\sum_{k}\left( C_{1k}|1_{1k}\rangle\prod_{k^{\prime}\neq k}|0_{{1k}^{\prime}}\rangle\prod_{k}{|0_{2k}\rangle} + C_{2k}|1_{2k}\rangle\prod_{k^{\prime}\neq k}|0_{{2k}^{\prime}}\rangle\prod_{k}{|0_{1k}\rangle}\right) +
%C_{c}|1_{c}\rangle{|\Psi}_{g}\rangle\prod_{k}{|0_{1k}\rangle}\prod_{k}{|0_{2k}\rangle}\nonumber\\ + C_{e}|0_{c}\rangle{|\Psi}_{e}\rangle\prod_{k}{|0_{1k}\rangle\prod_{k}{|0_{2k}\rangle}} 
%+ C_{0}|0_{c}\rangle{|\Psi}_{g}\rangle\prod_{k}{|0_{1k}\rangle\prod_{k}{|0_{2k}\rangle}}
%\end{eqnarray}
%\end{widetext}
\begin{widetext}
\begin{eqnarray}
|\Psi\rangle = &|0_{c}\rangle{|\Psi}_{g}\rangle\sum_{k}\Big(C_{1k}|1_{1k}\rangle\prod_{k^{\prime}\neq k}|0_{{1k}^{\prime}}\rangle\prod_{k}{|0_{2k}\rangle} + C_{2k}|1_{2k}\rangle\prod_{k^{\prime}\neq k}|0_{{2k}^{\prime}}\rangle\prod_{k}{|0_{1k}\rangle}\Big) +
C_{c}|1_{c}\rangle{|\Psi}_{g}\rangle\prod_{k}{|0_{1k}\rangle}\prod_{k}{|0_{2k}\rangle}\nonumber\\ &+ C_{e}|0_{c}\rangle{|\Psi}_{e}\rangle\prod_{k}{|0_{1k}\rangle\prod_{k}{|0_{2k}\rangle}} 
+ C_{0}|0_{c}\rangle{|\Psi}_{g}\rangle\prod_{k}{|0_{1k}\rangle\prod_{k}{|0_{2k}\rangle}},
\label{eqn29}
\end{eqnarray}
\end{widetext}
where $\Psi_{e} = \Psi_{e_{\pm}}$ depending on the symmetry of the field. We will later assume for definiteness that the cavity mode is coupled to the $\Psi_{e+} $ state. 
Next we use the SSE approach \cite{tokman2021,tokman2023}, which can be derived from both density-matrix and Heisenberg-Langevin formalisms \cite{plenio1998,scully,carmichael,meystre,zoller}. It gives equations
of evolution for the probability amplitudes $C_{\alpha}$ in a given
basis $|\alpha\rangle$, so that the equation of motion
of dyadics $\overline{C_{\alpha}C_{\alpha^{\prime}}^{\ast}}$ coincides with
the one for the density matrix elements $\rho_{\alpha\alpha^{\prime}}$ in
the Lindblad approximation. Here the bar
$\overline{\left(\ldots\right)}$ means averaging over the statistics
of Langevin noise sources which describe the effect of dissipative
reservoirs on the field and emitter degrees of freedom. Initial conditions for the SSE correspond to the initial values of the density matrix elements: $\overline{C_{\alpha}(0)C_{\alpha^{\prime}}^{\ast}(0)}=\rho_{\alpha\alpha^{\prime}}(0)$. Neglecting
field dissipation in the photon field outside the cavity, similarly to \cite{tokman2021,tokman2023}
we obtain the equations for the amplitudes of the state (\ref{eqn29}),
\begin{eqnarray}
\frac{dC_{1k}}{dt} + i\omega_{k}C_{1k} - i\mathfrak{L}^{\ast}C_{c} = 0\label{eqn30}\\
\frac{dC_{c}}{dt} + \left(i\omega_{c} + \frac{\mu_{c}}{2}\right)C_{c} - i\mathfrak{L}\sum_{k}C_{1k}\nonumber\\- i\Omega_{c}^{\ast}C_{e} = -\frac{i}{\hbar}\mathfrak{R}_{c}\left(t \right)\label{eqn31}\\
\frac{dC_{e}}{dt} + \left( i\frac{W_{e}}{\hbar} + \frac{\gamma}{2} \right)C_{e} - i\Omega_{c}C_{c} = - \frac{i}{\hbar}\mathfrak{R}_{e}\left( t \right)\label{eqn32}\\
\frac{dC_{2k}}{dt} + i\omega_{k}C_{2k} = 0\label{eqn33}\\
\frac{dC_{0}}{dt} = - \frac{i}{\hbar}\mathfrak{R}_{0}\left( t \right)\label{eqn34}.
\end{eqnarray}
Here we dropped the vacuum frequency shifts $\frac{\omega_{c,k}}{2}$,
which do not affect the observable quantities in this case.
In Eq.~(\ref{eqn31}) $\mu_{c}$ is the decay rate of the cavity mode
intensity due to the sum of Ohmic and diffraction losses. However,
$\mu_{c}$ does not include outcoupling into the modes of the outgoing
beam, since the latter effect is already included in Eq.~(\ref{eqn31})
via the term $\propto i\mathfrak{L}$. In Eq.~(\ref{eqn32})
$\gamma = \gamma_{e} + 2\gamma^{\left(el\right)}$, where
$\gamma_{e}$ is the relaxation rate of populations of QEs and
$\gamma^{\left(el\right)}$ is the elastic relaxation
rate, or pure dephasing. Eqs.~(\ref{eqn31}), (\ref{eqn32}), and (\ref{eqn34}) contain Langevin noise sources
$\mathfrak{R}_{\alpha}\left( t \right)$, where $\alpha = c,e,0$.
They satisfy
\begin{equation}
\overline{\mathfrak{R}_{\alpha}\left( t \right)} = 0,\
\overline{\mathfrak{R}_{\alpha}^{\ast}\left( t \right)\mathfrak{R}_{\alpha^{\prime}}\left( t^{\prime} \right)} = \hbar^{2}\delta\left( t - t^{\prime} \right)D_{\alpha\alpha^{\prime}}.
\label{eqn35}
\end{equation}

For low temperature of the reservoirs as compared to the optical
transition energy, the only nonzero correlators are \cite{tokman2021,tokman2023} 
\begin{equation}
D_{00} = \mu_{c}\overline{|C_{c}|^{2}} + \gamma_{e}\overline{|C_{e}|^{2}},\
D_{ee} = 2\gamma^{\left(el\right)}\overline{|C_{e}|^{2}},
\label{eqn36}
\end{equation}
which simplifies the solution. 
%\begin{eqnarray}
%\frac{dC_{1k}}{dt} + i\omega_{k}C_{1k} - i\mathfrak{L}^{\ast}C_{c} = 0 &&\label{eqn37} \\
%\frac{dC_{c}}{dt} + \left(i\omega_{c} + \frac{\mu_{c}}{2}\right)C_{c} - i\sum_{k}{\mathfrak{L}C_{1k}} - i\Omega_{c}^{\ast}C_{e} = 0&& \label{eqn38} \\
%\frac{dC_{e}}{dt} + \left(i\frac{W_{e}}{\hbar} + \frac{\gamma}{2}\right)C_{e} - i\Omega_{c}C_{c} = - \frac{i}{\hbar}\mathfrak{R}{e}\left(t\right)&& %\label{eqn39} \\
%\frac{dC{2k}}{dt} + i\omega_{k}C_{2k} = 0 &&\label{eqn40} \\
%\overline{C_{0}\left(t\right) - C_{0}\left(0\right)} = 0,&& \nonumber \\
%\overline{\left|C_{0}\right|^{2} + \sum_{k}\left(\left|C_{1k}\right|^{2} + \left|C_{2k}\right|^{2}\right) + \left|C_{c}\right|^{2} + \left|C_{e}%\right|^{2}} &=&1\nonumber\label{eqn41}.\\
%\end{eqnarray}
%\begin{eqnarray}
%\frac{dC_{1k}}{dt} + i\omega_{k}C_{1k} - i\mathfrak{L}^{\ast}C_{c} = 0\label{eqn37}\\
%\frac{dC_{c}}{dt} + \left(i\omega_{c} + \frac{\mu_{c}}{2}\right)C_{c} - i\sum_{k}{\mathfrak{L}C_{1k}} - i\Omega_{c}^{\ast}C_{e} = %0\label{eqn38}\\
%\frac{dC_{e}}{dt} + \left(i\frac{W_{e}}{\hbar} + \frac{\gamma}{2}\right)C_{e} - i\Omega_{c}C_{c} = - \frac{i}{\hbar}\mathfrak{R}_{e}\left(t\right)\label{eqn39}\\
%\frac{dC_{2k}}{dt} + i\omega_{k}C_{2k} = 0\label{eqn40}\\
%\overline{C_{0}\left(t\right) - C_{0}\left(0\right)} = 0,\nonumber\\
%\overline{\left|C_{0}\right|^{2} + \sum_{k}\left(\left|C_{1k}\right|^{2} + \left|C_{2k}\right|^{2}\right) + \left|C_{c}\right|^{2} + \left|C_{e}\right|^{2}} = 1\label{eqn41}.
%\end{eqnarray}
Note that the above equations for state amplitudes conserve the noise-averaged norm of the state
vector, $\overline{\left|C_{0}\right|^{2} + \sum_{k}\left(\left|C_{1k}\right|^{2} + \left|C_{2k}\right|^{2}\right) + \left|C_{c}\right|^{2} + \left|C_{e}\right|^{2}} = 1$. Furthermore, Eq.~(\ref{eqn33}) for $C_{2k}$ is decoupled from the other
equations. If we neglect pure dephasing one can put
$\mathfrak{R}_{e}=0$ in Eq.~(\ref{eqn32}). For a finite value of
$\gamma^{\left(el\right)}$ Eqs.~(\ref{eqn35}) and (\ref{eqn36}) yield
\begin{equation}
\overline{\mathfrak{R}_{e}^{\ast}\left(t\right)\mathfrak{R}_{e}\left(t^{\prime}\right)} = 2\gamma^{\left(el\right)}\hbar^{2}\delta(t - t^{\prime})\overline{\left|C_{e}\right|^{2}}.
\label{eqn42}
\end{equation}

As an initial condition, we consider a single-photon pulse incident on
the cavity which contained no photons, i.e.,
\begin{eqnarray}
C_{c}\left(0\right) = C_{e}\left(0\right) = C_{0}\left(0\right) = 0,\label{eqn43}\\
\sum_{k = k_{\min}}^{k_{\max}}{\left(|C_{1k}\left(0\right)|^{2} + |C_{2k}\left(0\right)|^{2}\right) = 1}.
\label{eqn44}
\end{eqnarray}
Initial conditions~(\ref{eqn43}, \ref{eqn44}) imply that at the initial
moment of time \emph{t} = 0 a single-photon pulse has not reached the
cavity yet. The pulse position at \emph{t} = 0 is determined by a set of
initial phases $\varphi_{\left(1,2\right)k} = {\rm Arg}\left\lbrack C_{\left(1,2\right)k}\left(0\right) \right\rbrack$.
In Appendix A we show that if a pulse with spatial length
$\sim l_{p}$ is located at $t = 0$ around the cross section
$s = s_{0}$, where $s_{0} < 0$, $|s_{0}|\gg l_p$, the corresponding phases ensure that
\begin{equation}
\left|\sum_{k}{C_{\left(1,2\right)k}\left(0\right)e^{-i\omega_{k}t}}\right|_{\begin{matrix}
t < T_{0} \\
t > T_{1} \\
\end{matrix}}
\rightarrow 0
\label{eqn45}
\end{equation}
where
$T_{0,1}\approx\frac{\left|s_{0}\right|\mp l_{p}}{\upsilon_{g}}$,
$\upsilon_{g}$ is the group velocity. In order to derive the final solution for the state vector we will use both
Eq.~(\ref{eqn45}) and the result which follows from it at $ t = 0$:
\begin{equation}
\left|\sum_{k}{C_{\left(1,2\right)k}\left(0\right)}\right|\rightarrow 0
\label{eqn46}
\end{equation}
Eqs.~(\ref{eqn45}) and~(\ref{eqn46}) have a simple interpretation: at $t < T_{0}$
(which includes $t = 0$) the incident pulse has not reached the cross
section $s = 0$ yet, whereas at $t > T_{1}$ it has already left this
cross section. 

The details of solving the evolution equations for complex amplitudes are in Appendix B.    
The resulting solution for the state vector of the system within the SSE formalism at times
$t > T_{1}$ is given by
\begin{widetext}
\begin{eqnarray}
\left|\Psi\left(t > T_{1}\right)\right\rangle
\approx &\left|\Psi_{g} \right\rangle\left|0_{c}\right\rangle \times \left\{\sum_{k}{e^{- i\omega_{k}t}\left\lbrack R_1 C_{1k}\left(0 \right)\left|1_{1k}\right\rangle\prod_{k^{\prime}\neq k}\left|0_{{1k}^{\prime}} \right\rangle\prod_{k}{|0_{2k}\rangle} + C_{2k}\left(0 \right)\left|1_{2k}\right\rangle\prod_{k^{\prime}\neq k}\left|0_{{2k}^{\prime}}\right\rangle\prod_{k}{|0_{1k}\rangle}\right\rbrack}\right.\nonumber\\
&\left.+ 
%\sum_{k}{\delta C_{1k}|1_{1k}\rangle\prod_{k^{\prime}\neq k}|0_{{1k}^{\prime}}\rangle\prod_{k}{|0_{2k}\rangle}} 
C_{0}\prod_{k}|0_{1k}\rangle\prod_{k}{|0_{2k}\rangle}\right\}.
\label{eqn68}
\end{eqnarray}
\end{widetext}
Here
\begin{equation}
R_1 = 1 - \frac{2\left(p_{e}-i\Delta_{0}\right)\kappa}{\left(P_{1}+i\Delta_{0}\right)\left(P_{2} + i\Delta_{0}\right)}
\label{eqn69}
\end{equation}
is the crucial parameter of the solution which determines the probability amplitude of the reflected quantum field in the direction parallel to the cavity polarization axis $e_1$: 
\begin{eqnarray}
C_{1k}\left(t > T_{1}\right) &=& R_1C_{1k}\left( 0 \right)e^{- i\omega_{k}t} \label{eqn75a} \\
C_{2k}\left( t > T_{1}\right) &=& C_{2k}(0) e^{- i\omega_{k}t}. \label{eqn75b} 
\end{eqnarray}
Here Eq.~(\ref{eqn75b}) means simply that the photon polarization along $e_2$ axis does not couple to the cavity mode and is reflected with reflectivity 1. 
The derivation of $R_1$ is nontrivial and is detailed in Appendix B. The closed form (\ref{eqn69}) is obtained for a narrowband incident pulse and neglecting pure dephasing; the impact of the latter is discussed in Appendix C. The geometric and relaxation parameters entering Eq.~(\ref{eqn69}) are as follows: $p_{e} = \frac{\gamma}{2} + i\Delta_{e}$, where $\Delta_e = \frac{W_{e+}}{\hbar} - \omega_c$ is the detuning between the bright-state transition frequency and the cavity mode frequency; 
\begin{equation}
P_{1,2} = - \frac{\kappa_{\Sigma} + p_{e}}{2} \pm \sqrt{\left( \frac{\kappa_{\Sigma} - p_{e}}{2} \right)^{2} - \left| \Omega_{c} \right|^{2}},
\label{eqn59-1}
\end{equation}
$\kappa_{\Sigma} = \frac{\mu_{c}}{2} + \kappa$ is the total rate of cavity mode losses, which includes the sum of the Ohmic absorption and scattering losses $\mu_c$ and the rate of in-and out-coupling of an incident photon pulse to the cavity,  
$\kappa$, which should be calculated separately, by solving the classical electrodynamic problem for a given geometry; see the details in Appendix B. Finally, $\Delta_{0} = \omega_0 - \omega_c$ is the detuning between the central frequency of the incident single-photon wavepacket and the cavity mode frequency.

%%%%%%%%%%%%%%%%%%%%%%%%%%%%%%%%%%%%%%%%%%%%%%%%%%%%%%%%%%%%%%%%%%%%%%%%%%%%%%%%%%
%%%%%%%%%%%%%%%%%%%%%%%%%%%%%%%%%%%%%%%%%%%%%%%%%%%%%%%%%%%%%%%%%%%%%%%%%%%%%%%%%%
\section{Control of the polarization state of a single photon upon reflection from the cavity}
%%%%%%%%%%%%%%%%%%%%%%%%%%%%%%%%%%%%%%%%%%%%%%%%%%%%%%%%%%%%%%%%%%%%%%%%%%%%%%%%%%%

From  Eqs.~(\ref{eqn68}) and~(\ref{eqn69}), one can calculate the polarization quantum state of the reflected photon for given parameters of the system, the state of the QEs prepared initially by the classical field, and the quantum state of the incident photon. We will consider the incident photon to have a linear polarization. For arbitrary detunings the factor $R_1$ is complex-valued and the reflected polarization will be elliptical. For applications, it is more convenient to deal with  a fully resonant case, when $R_1$ is real and the reflected polarization remains linear.

\begin{figure}[h!]
\centering
\includegraphics[width=8.5cm,height=4.6cm]{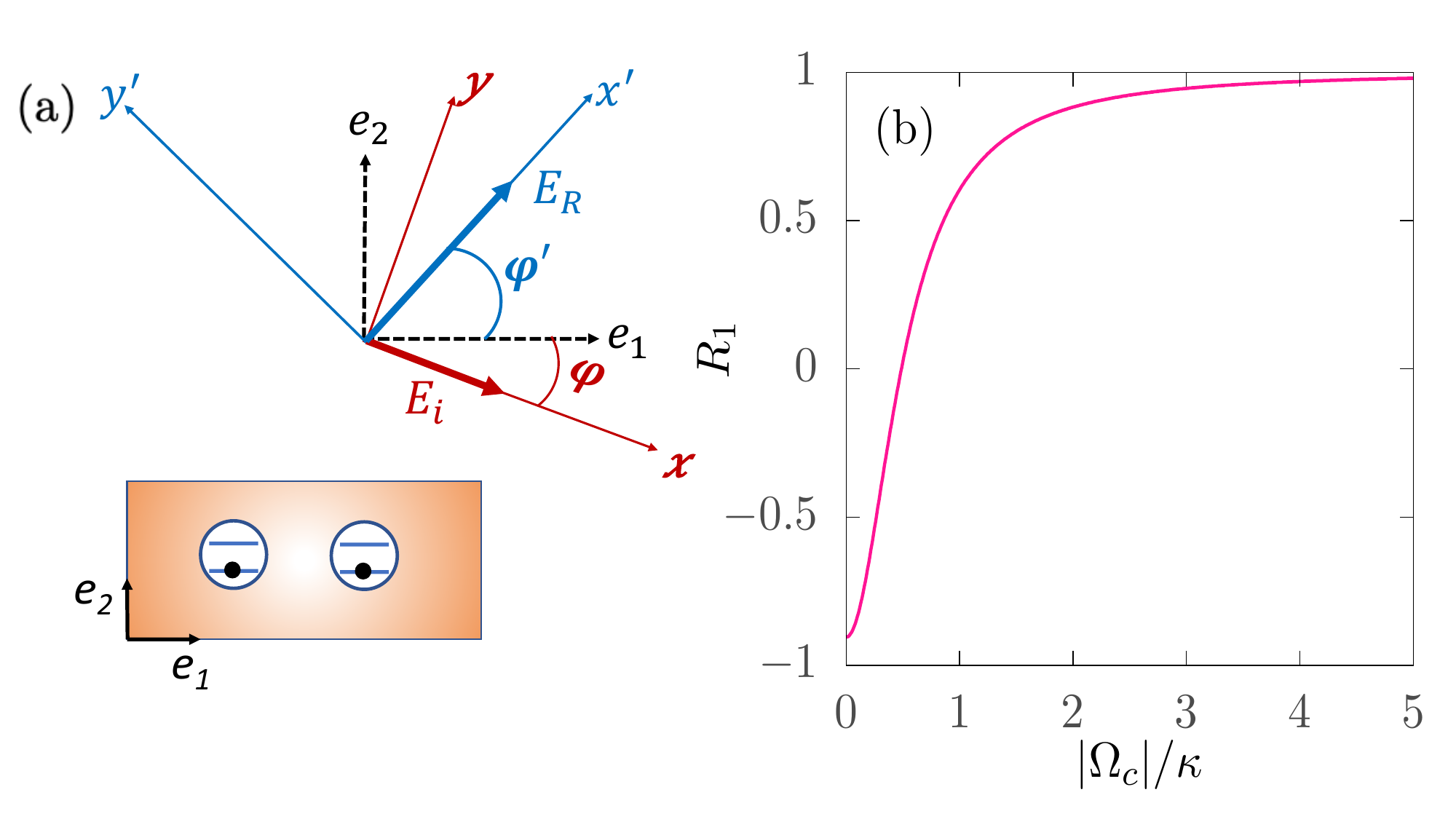}
\vspace{-0.1cm}
\caption{Rotation of photon polarization upon reflection from the cavity. (a) The cavity axes $(e_1,e_2)$ with respect to the orientation of coordinate systems $(x,y)$  and $(x',y')$ in which the polarizations of incident and reflected photons (respectively) are measured. $E_i$ and $E_R$ are electric fields of the incident and reflected photon. (b) The reflection coefficient $R_1$ from Eq.~(\ref{eqn69res}) as a function of the Rabi frequency $|\Omega_{c}|/\kappa$ for $\gamma/\kappa=0.5$ and $\mu_{c}/\kappa=0.1$.}
\label{fig2}
\end{figure}

Consider only one mode of the ray bundle, i.e., drop the index
 $k$. When $t > T_{1}$ the excitations of the cavity
mode and QEs have already decayed, so we only need the wave function of
the external field. Let the polarization of the incident and reflected photon be measured in
the $(x,y)$ coordinate system, whereas the cavity has
axes $(e_1,e_2)$ as shown in Fig.~2(a).  The transformation rules for the polarization states in different bases are derived in Appendix D. Using these rules, one can show
that if the wave function in the coordinates $(x,y)$ is defined by
\begin{equation}
\left|\Psi^{\left(x,y \right)}\right\rangle = \frac{A|1_{x}\rangle|0_{y}\rangle + B|0_{x}\rangle|1_{y}\rangle + C|0_{x}\rangle|0_{y}\rangle}{\sqrt{\left|A\right|^{2} + \left|B\right|^{2}+{\left|C\right|}^{2}}},
\label{eqn75}
\end{equation}
then in the coordinates $(e_1,e_2)$ it takes the form
\begin{widetext}
\begin{equation}
\left|\Psi^{\left(e_1,e_2\right)}\right\rangle = \frac{\left(A\cos\phi + B\sin\phi\right)|1_{e_1}\rangle|0_{e_2}\rangle + \left(B\cos\phi - A\sin\phi\right)|0_{e_1}\rangle|1_{e_2}\rangle + C|0_{e_1}\rangle|0_{e_2}\rangle}{\sqrt{\left|A\right|^{2} + \left|B\right|^{2} + {\left|C\right|}^{2}}}
\label{eqn78}
\end{equation}
\end{widetext} 
where $\phi$ is the rotation angle of the system $(e_1,e_2)$
with respect to (x,y) (see Fig.~\ref{fig2}).

%%%%%%%%%%%%%%%%%%%%%%%%%%%%%%%%%%%%%%%%%%%%%%%%%%%%%%%%%%%%%%%%%%%%%%%%%%%%%%%%%%%%
\label{section4}

%%%%%%%%%%%%%%%%%%%%%%%%%%%%%%%%%%%%%%%%%%%%%%%%%%%%%%%%%%%%%

Now consider the case
of the incident photon polarized along x,
\begin{equation}
|\Psi^{\left( x,y \right)}\rangle_{inc} = |1_{x}\rangle|0_{y}\rangle.
\label{eqn79}
\end{equation}
Then  in the cavity axes coordinates $(e_1,e_2)$ the quantum state of this photon is 
\begin{equation}
|\Psi^{\left(e_1,e_2\right)}\rangle_{inc} = \cos\phi|1_{e_1}\rangle|0_{e_2}\rangle - \sin\phi|0_{e_1}\rangle|1_{e_2}\rangle.
\label{eqn80}
\end{equation}
The photon quantum state after reflection can be obtained using Eqs.~(\ref{eqn75a}) and (\ref{eqn75b}):
\begin{eqnarray}
|\Psi^{\left( e_1,e_2\right)}\rangle_{ref} =
R_{1}\cos\phi|1_{e_1}\rangle|0_{e_2}\rangle - \sin\phi|0_{e_1}\rangle|1_{e_2}\rangle\nonumber\\ + \sqrt{1-\left(R_{1}\cos\phi\right)^2 - \left(\sin\phi\right)^{2}}|0_{e_1}\rangle|0_{e_2}\rangle.
\label{81}
\end{eqnarray}
%where, in agreement with Eq.~(\ref{eqn41}), $\overline{\zeta}=0$, $\overline{|\zeta|^{2}}=1$.

To interpret this result, it is convenient to introduce the coordinate
system $(x',y')$ in which the reflected electric field $E_R$ is along $x'$ axis, i.e.,  the reflected photon quantum state is 
\begin{equation}
|\Psi^{\left(x',y'\right)}\rangle_{ref} = C_{x'}|1_{x'}\rangle|0_{y'}\rangle
+C_{0'}|0_{x'}\rangle|0_{y'}\rangle,
\label{82}
\end{equation}
where $\overline{C_{x'}^{\ast}C_{0'}}$=0, $\overline{|C_{x',0'}|^{2}}\neq0$. Using the transformation (\ref{eqn78}),
we obtain that the coordinate system
$\left(x',y'\right)$ is rotated with respect to the cavity axes $(e_1,e_2)$ by an angle $\phi^{\prime}$ defined by
\begin{equation}
-\frac{1}{R_{1}}\tan\phi = \tan\phi^{\prime}
\end{equation}

Now we consider various limiting cases of this result. To keep the value of $R_1$ real and the reflected polarization linear, we need to take all detunings equal to zero, which gives 
\begin{equation}
R_1 = 1 - \frac{\gamma \kappa}{\frac{\gamma}{2}\left(\kappa + \frac{\mu_c}{2} \right)+ |\Omega_c|^2}. 
\label{eqn69res}
\end{equation}
The plot of $R_1$ in Eq.~(\ref{eqn69res}) as a function of $|\Omega_{c}|/\kappa$ is shown in Fig.~2(b). 
If the QEs are in the ground state or the bright state and are strongly coupled to a cavity mode, i.e., $|\Omega_{c}|> \mu_{c}, \gamma,\kappa$, 
we have $R_1 \approx 1$, in which case $\phi^{\prime}=-\phi$, i.e., the photon polarization does not change upon reflection. 

When the QEs are prepared in the dark state decoupled from the cavity mode, one can put $|\Omega_{c}|^2 \approx 0$ in Eq.~(\ref{eqn69res}). In this case $R_1 \rightarrow 1$ for $\frac{\mu_{c}}{2\kappa}\gg1$ and $R_1 \rightarrow -1$ for $\frac{\mu_{c}}{2\kappa} \ll 1$. Assuming the latter case, we obtain $\phi^{\prime}=\phi$ i.e., the axis $e_1$ of the cavity bisects the polarization directions of the incident and reflected fields. In particular, if $\phi=\pi/4$, the photon polarization is rotated by $\pi/2$ upon reflection. This opens up a possibility of a two-qubit gate with the cavity being a control qubit  and the polarization state of an incident photon being a flying signal qubit. The cavity reflection is controlled by preparing the state of QEs with an external  classical field. 

More complicated control scenarios are possible depending on the magnitude and phase of the reflection factor in Eq.~(\ref{eqn69}). In particular,
one can choose the initial conditions so that  $R_1=0$, which corresponds to the critical coupling regime \cite{yariv}. In this case the reflected field is polarized along the $e_2$ axis of the cavity.

The critical coupling is possible even at finite frequency detuning
$\Delta_{0}$ between the incident photon and the cavity mode. Consider for simplicity an exact resonance between the
cavity field and QEs, when $\Delta_{e} = \frac{W_{e}}{\hbar} - \omega_{c} = 0$.
In this case Eq.~(\ref{eqn75a}) becomes
\begin{equation}
C_{1k}\left( t > T_{1} \right) = C_{1k}\left( 0 \right)\left(1 - \frac{2\left( \frac{\gamma}{2}  - i\Delta_{0} \right)\kappa}{D} \right)e^{- i\omega_{k}t},
\label{eqne1}
\end{equation}
where
\begin{eqnarray}
D &=&- \left(\Delta_{0} - iP_{1} \right)\left( \Delta_{0} - iP_{2} \right)\nonumber\\&&=\left(\frac{\gamma}{2}  - i\Delta_{0} \right)\left(\kappa + \frac{\mu_{c}}{2} - i\Delta_{0} \right) + \left|\Omega_{c}\right|^{2}.
\label{eqne2}
\end{eqnarray}
The factor $\frac{1}{D}$ describes the resonant lineshape of
the cavity split due to the coupling to the QEs. The critical coupling regime corresponds to the condition
\begin{equation}
\left(\frac{\gamma}{2} - i\Delta_{0} \right)\left(\frac{\mu_{c}}{2} - i\Delta_{0} \right) + \left|\Omega_{c}\right|^{2} = \left(\frac{\gamma}{2} - i\Delta_{0}\right)\kappa,
\label{eqne3}
\end{equation}
Which can be satisfied when
\begin{equation}
\kappa = \frac{\mu_{c}}{2} + \frac{\gamma}{2},\
\Delta_{0}^{2} + \frac{\gamma^{2}}{4} = \left|\Omega_{c}\right|^{2};
\label{eqne4}
\end{equation}
or
\begin{equation}
\kappa = \frac{\mu_{c}}{2} + \frac{2\left|\Omega_{c}\right|^{2}}{\gamma},\
\Delta_{0} = 0 
\label{eqne5}
\end{equation}
In the strong coupling regime, when
$\left|\Omega_{c}\right| \gg \kappa,\frac{\mu_{c}}{2},\gamma$,
critical coupling is possible in the case (\ref{eqne4}). In
this case the field frequency is resonant to one of the maxima of the
cavity lineshape with accuracy up to the terms
$\sim\frac{\left(\kappa + \frac{\mu_{c}}{2} - \frac{\gamma}{2}\right)^{2}}{{8\left|\Omega_{c}\right|}^{2}}$. 

Finally, even more opportunities for controlling the quantum state of a reflected photon are created when the QEs are prepared in an arbitrary superposition of bright and dark states. Since the derivation is cumbersome, we present it in Appendix E below.

%%%%%%%%%%%%%%%%%%%%%%%%%%

\section{Discussion and conclusions}
\label{section5}
%%%%%%%%%%%%%%%%%%%%%%%%%%
%%%%%%%%%%%%%%%%%%%%%%%%%%

Here we discuss feasibility and constraints on possible implementation
of this approach to quantum state control and gating. First of all, one needs an anisotropic, polarization-selective nanocavity. For a solid-state
dielectric nanocavity one could use an anisotropic 2D photonic crystal
cavity, similar to the one utilized in \cite{waks2013,waks2018} or a Fabry-Perot or DBR cavity with a highly elongated cross-section. Within the plasmonic
platform there are many possible anisotropic cavity architectures, e.g.
a bow-tie cavity \cite{bitton2019}, metal nanorods or antennas etc. \cite{dovzhenko2018}. Next, for the particular implementation discussed above, one needs different spatial symmetries of the quantized cavity mode and the classical field used in excitation of the QEs into the dark state. The easiest way is to
couple the QEs to a quantized symmetric mode of a cavity at frequency
$\omega_{s}$, with quantum emitters located in the middle of the
cavity near the maximum of the field distribution. If the vacuum Rabi
frequency for this mode $\Omega_{c}$, the strong coupling condition is
$\Omega_{c}\gg\gamma$, where $\gamma$ is the relaxation constant
for QEs. The (strong) classical field which transfers the QEs into the
dark state has to be centered at frequency $\omega_{a} = \omega_{s} - 2\Omega_{dd}$.
It is important that the spatial distribution of the classical field is asymmetric and has a
large field gradient at the position of QEs. 
This will maximize the coupling to the antisymmetric dark state of QEs,
which is described by the classical Rabi frequency
$\Omega_{a}^{\left(cl\right)}$. The pulse of a classical
field should have an area close to $\pi$, i.e.,
$\Omega_{a}^{\left(cl\right)}t\sim\pi$, where t is a
pulse duration and one needs $t^{- 1} > \gamma$. The classical field
should not couple to a symmetric bright state of QEs, which implies
$\Omega_{s}^{\left(cl\right)} < 2\Omega_{dd}$.

If the coupling of the classical field to a single QE is characterized
by classical Rabi frequency $\sim\Omega^{\left(cl\right)}$,
for the antisymmetric Rabi frequency one can write
$\Omega_{a}^{\left(cl\right)}\sim\frac{\mathrm{\Delta}}{L}\Omega^{\left(cl\right)}$,
where $\mathrm{\Delta}$ is the distance between QEs and L is the
scale of the spatial nonuniformity of the classical electric field. At
the same time, the Rabi frequency which characterizes the coupling of
the classical field to the symmetric state of the QEs can be
parametrized as $\Omega_{s}^{\left(cl\right)}\sim\alpha\Omega^{\left(cl\right)}$,
where the coefficient $0\leq\alpha < 1$ is determined by the
position of the node of the classical field with respect to QEs. All of
the above restrictions can be expressed through the constraint of the
effective number of photons in the classical pulse,
$\sim\frac{\left(\Omega^{\left(cl\right)}\right)^{2}}{\Omega_{c}^{2}}$:
\begin{equation}
\frac{1}{\alpha}\frac{\Omega_{dd}}{\Omega_{c}} > \sqrt{n} > \frac{L}{\mathrm{\Delta}}\frac{\gamma}{\Omega_{c}}.
\end{equation}
To maximize the dynamic range of possible classical pulse intensities, one needs a small $\alpha$, a large enough dipole-dipole splitting $\Omega_{dd}$, small $\gamma < \Omega_c$ (i.e., a strong coupling condition), and a large classical field gradient $\propto 1/L$. The latter could be maximized by illuminating a tip or nanoantenna on top of the cavity for better control of the excitation field distribution. As an example, the optical dipole-dipole interaction energy for semiconductor quantum dots at $\sim 10$ nm from each other is $\sim 1$ meV, whereas the vacuum Rabi frequency in a photonic crystal cavity is of the order of 100 $\mu$eV. The value of $\gamma$ varies greatly, but cryogenic cooling is usually required to reach strong coupling regime in dielectric cavities  \cite{lodahl2015,dovzhenko2018,deppe,reithmaier}. 

In conclusion, we proposed and analyzed quantum gate operation based on single photons interacting with a nanocavity loaded with quantum emitters such as quantum dots or molecules. Exciting QEs into an entangled dark state changes the polarization state of a reflected photon. A variety of different dynamic regimes and operations is possible. We developed a formalism based on the stochastic Schroedinger equation which includes relaxation and dephasing processes for all degrees of freedom. Moreover, it allowed us to obtain analytic solutions for the quantum state of the reflected photon depending on the state of QEs and relaxation and coupling parameters. 

%%%%%%%%%%%%%%%%%%%%%%%%%%%%%%%%%%%%%%%%%%%%

\begin{acknowledgments}
This work has been supported in part by the Air Force Office for Scientific Research 
Grant No.~FA9550-21-1-0272, National Science Foundation Award No.~1936276, and Texas A\&M University through the STRP program. M. T. acknowledges the support by the Center for Integration in Science of the Ministry of Aliya and Integration, Israel.

\end{acknowledgments}

%%%%%%%%%%%%%%%%%%%%%%%%%%%%%%%

%%%%%%%%%%%%%%%%%%%%%%%%%
\appendix

%%%%%%%%%%%%%%%%%%%%%%%%%%%%%%%%%%%%%%%%%%%%%%%%%%%%%%

\section{Propagation of the single-photon optical pulse}

%%%%%%%%%%%%%%%%%%%%%%%%%%%%%%%%%%%%%%%%%%%%%%%%%%%%%%%
%%%%%%%%%%%%%%%%%%%%%%%%%%%%%%%%%%%%%%%%%%%%%%%%%%%%%%%
Equations (\ref{eqn45}) and (\ref{eqn46}) in the main text can be obtained from the analysis of
propagation of a single-photon pulse of finite length and cross section (the ray bundle, as sketched in Fig.1). Here we find the space-time field intensity profile in such a pulse. Since
we assume degenerate eigenmodes of a ray bundle, the results for the
eigenmodes with orthogonal polarizations $\mathbf{e}_{1}$ and
$\mathbf{e}_{2}$ are identical:
%\begin{widetext}
\begin{eqnarray}
I_{1,2}(s,t) &=& \Big\langle\Psi(t)\Big|\Bigg\{\sum_{k=k_{\min}}^{k_{\max}}\Bigg[ E_{k}(\xi_{\bot}){\hat{c}}_{(1,2)k}e^{iks}\nonumber\\
&&+E_{k}^{\ast}(\xi_{\bot}){\hat{c}}_{(1,2)k}^{\dagger}e^{-iks}\Bigg]\Bigg\}^{2}\Big|\Psi(t)\Big\rangle.
\label{eqna1}
\end{eqnarray}
Consider the EM field of a pulse which has not reached the cavity yet.
In this case the state of QEs in a cavity is unimportant, and one can put
%\begin{eqnarray}
%\left|\Psi\left( t \right)\right\rangle = &&
%\sum_{k=k_{\min}}^{k_{\max}}{e^{- i\omega_{k}t}\Bigr(C_{1k}\left( 0 \right)|1_{1k}\rangle\prod_{k^{\prime}\neq k}|0_{{1k}^{\prime}}\rangle\prod_{k}{|0_{2k}\rangle} +\\&& C_{2k}\left( 0 \right)|1_{2k}\rangle\prod_{k^{\prime}\neq k}\left|0_{{2k}^{\prime}}\right\rangle\prod_{k}{|0_{1k}\rangle}\Bigr)}.
%\label{eqna2}
%\end{eqnarray}
\begin{eqnarray}
\left|\Psi\left(t\right)\right\rangle &=& \sum_{k=k_{\min}}^{k_{\max}} e^{-i\omega_{k}t} \Bigg( C_{1k}\left(0\right)|1_{1k}\rangle\prod_{k^{\prime}\neq k}|0_{{1k}^{\prime}}\rangle\prod_{k} |0_{2k}\rangle \nonumber \\
&& + C_{2k}\left(0\right)|1_{2k}\rangle\prod_{k^{\prime}\neq k}|0_{{2k}^{\prime}}\rangle\prod_{k} |0_{1k}\rangle \Bigg).
\label{eqna2}
\end{eqnarray}
For a narrow spectrum we can take
$\mathbf{E}_{k} \approx \mathbf{E}_{0}$. All averages of the dyadics
${\hat{c}}_{\left(1,2\right)k}{\hat{c}}_{\left(1,2\right)q}$
and
${\hat{c}}_{\left(1,2\right)k}^{\dagger}{\hat{c}}_{\left(1,2\right)q}^{\dagger}$
are zero, which gives
\begin{equation}
I_{1,2} = \left|E_{0}\left(\mathbf{\xi}_{\mathbf{\bot}} \right)\right|^{2}\langle\Psi\left(0\right) |\hat{T}| \Psi\left( 0 \right)\rangle,
\label{eqna3}
\end{equation}
where
\begin{eqnarray}
\hat{T} = \sum_{k = k_{\min}}^{k_{\max}}\sum_{q = k_{\min}}^{k_{\max}} & \Bigl[{\hat{c}}_{(1,2)k}{\hat{c}}_{(1,2)q}^{\dagger}e^{-i(\omega_{k} - \omega_{q})t + i(k - q)s}\nonumber\\
 + &{\hat{c}}_{(1,2)k}^{\dagger}{\hat{c}}_{(1,2)q}e^{i(\omega_{k} - \omega_{q})t - i(k - q)s}\Bigr]
\label{eqnarray}
\end{eqnarray}

For a narrow spectrum we get
\begin{equation}
e^{- i\left( \omega_{k} - \omega_{q} \right)t + i\left( k - q \right)s} \Rightarrow e^{i\left( s - \upsilon_{g}t \right)\left( k - q \right)},
\label{eqna5}
\end{equation}
where \(\upsilon_{g} = \frac{\partial\omega_{k}}{\partial k}\) is a group velocity at the center of the pulse spectrum.
Taking into account that ${\hat{c}}_{\left(1,2\right)k}{\hat{c}}_{\left(1,2\right)q}^{\dagger} = {\hat{c}}_{\left( 1,2 \right)q}^{\dagger}{\hat{c}}_{\left( 1,2 \right)k} + \delta_{kq}$
and using Eq.~(\ref{eqna5}), we obtain
%\begin{equation}
%I_{1,2} = \left| E_{0}\left( \mathbf{\xi}_{\mathbf{\bot}} \right) \right|^{2}\left\{\frac{\mathrm{\Delta}kL}{2\pi} + 2\left|\sum_{k = k_{\min}}^{k_{\max}}{C_{\left(1,2 \right)k}\left(0\right)e^{i\left(s - \upsilon_{g}t\right)k}}\right|^{2}\right\},
%\label{eqna6}
%\end{equation}
%\begin{eqnarray}
%I_{1,2} = \left| E_{0}\left( \mathbf{\xi}_{\mathbf{\bot}} \right) \right|^{2}\Biggr\{\frac{\mathrm{\Delta}kL}{2\pi} + 2\\\left|\sum_{k = k_{\min}}^{k_{\max}}{C_{\left(1,2 \right)k}\left(0\right)e^{i\left(s - \upsilon_{g}t\right)k}}\right|^{2}\Biggr\},
%\label{eqna6}
%\end{eqnarray}
\begin{eqnarray}
I_{1,2} = \left|E_{0}\left( \mathbf{\xi}_{\mathbf{\bot}} \right) \right|^{2}\frac{\mathrm{\Delta}kL}{2\pi} + 2\left|E_{0}\left( \mathbf{\xi}_{\mathbf{\bot}} \right) \right|^{2}\times\nonumber\\\left|\sum_{k = k_{\min}}^{k_{\max}}{C_{\left(1,2 \right)k}\left(0\right)e^{i\left(s - \upsilon_{g}t\right)k}}\right|^{2},
\label{eqna6}
\end{eqnarray}
where $\mathrm{\Delta}k = k_{\max} - k_{\min}$,
$\frac{\mathrm{\Delta}kL}{2\pi} \approx \frac{\mathrm{\Delta}\omega L}{2\pi\upsilon_{g}}$
is the number of quantized field modes within the pulse bandwidth
$\mathrm{\Delta}\omega$. The first term in parentheses determines the
intensity of vacuum fluctuations. In the limit of a continuous spectrum,
when $\sum_{k = k_{1}}^{k_{2}}\left( \ldots \right) \Longrightarrow \frac{L}{2\pi}\int_{k_{1}}^{k_{2}}{\left( \ldots \right)\text{dk}}$,
%\begin{eqnarray}
%I_{1,2} = \left|E_{0}\left(\mathbf{\xi}_{\mathbf{\bot}}\right)\right|^{2}\Biggr\{\frac{\mathrm{\Delta}kL}{2\pi} + 2\left(\frac{L}{2\pi}\right)^{2}\times\nonumber\\\left|\int_{k_{\min}}^{k_{\max}}{C_{\left( 1,2 %\right)k}\left( 0 \right)e^{i\left( s - \upsilon_{g}t \right)k}dk}\right|^{2} \Biggr\}
%\label{eqna7}
%\end{eqnarray}
\begin{eqnarray}
I_{1,2} = \left|E_{0}\left(\mathbf{\xi}_{\mathbf{\bot}}\right)\right|^{2}\frac{\mathrm{\Delta}kL}{2\pi} + 2\left|E_{0}\left(\mathbf{\xi}_{\mathbf{\bot}}\right)\right|^{2}\left(\frac{L}{2\pi}\right)^{2}\times\nonumber\\\left|\int_{k_{\min}}^{k_{\max}}{C_{\left( 1,2 \right)k}\left( 0 \right)e^{i\left( s - \upsilon_{g}t \right)k}dk}\right|^{2}
\label{eqna7}
\end{eqnarray}
In this limit the normalization given by Eq.~(\ref{eqn44}) becomes
\begin{equation}
\frac{L}{2\pi}\int_{k = k_{\min}}^{k_{\max}}{\left( \left| C_{1k}\left( 0 \right) \right|^{2} + \left| C_{2k}\left( 0 \right) \right|^{2} \right)\text{dk}} = 1.
\label{eqna8}
\end{equation}
Since from Eq.~(\ref{eqn26}) we have the scaling
$E_{0}^{2} \propto \frac{1}{L}$, it is easy to see that the functions
$I_{1,2}$ given by Eq.~(\ref{eqna7}) do not depend on the quantization length
\emph{L}.
The field intensity (\ref{eqna7}) corresponds to the energy flux density
\begin{equation}
P_{1,2} = n_{0} c \frac{E_{0}^{2}\left( \mathbf{\xi}_{\mathbf{\bot}} \right)}{2\pi}\left( \frac{L}{2\pi} \right)^{2}\left| \int_{k_{\min}}^{k_{\max}}{C_{\left( 1,2 \right)k}\left( 0 \right)e^{i\left( s - \upsilon_{g}t \right)k}\text{dk}} \right|^{2},
\label{eqna9}
\end{equation}
where $n_{0}$ is the refractive index at the center of the spectral bandwidth.

The position of the pulse at the initial moment is given by initial
phases
$\varphi_{\left(1,2\right)k} = Arg\left\lbrack C_{\left(1,2\right)k}\left(0\right)\right\rbrack$.
For the pulse of spatial length
$l_{p} \approx \ \frac{2\pi}{\mathrm{\Delta}k}$ to be located at the
cross section $s = s_{0}$ at $t = 0$ the following conditions must
be met:

a) At $\left| s - s_{0} \right| < \frac{2\pi}{\mathrm{\Delta}k}$ in
the interval $k \in \left\lbrack k_{\min},k_{\max} \right\rbrack$
there must be at least one stationary phase point satisfying
\begin{equation}
\frac{\partial\varphi_{\left( 1,2 \right)k}}{\partial k} + s = 0.
\label{eqna10}
\end{equation}
b) At $\left| s - s_{0} \right| > \frac{2\pi}{\mathrm{\Delta}k}$ there
should be no stationary points. In this case the integral on the
right-hand side of Eq.~(\ref{eqna9}) approaches zero because of a rapidly
oscillating phase factor $e^{i(\varphi_{\left( 1,2 \right)k} + sk)}$. This gives the
conditions (\ref{eqn45}) and (\ref{eqn46}).

%%%%%%%%%%%%%%%%%%%%%%%%%%%%%%%%%%%%%%%%%%%%%%%
%%%%%%%%%%%%%%%%%%%%%%%%%%%%%%%%%%%%%%%%%%%%%%%

\section{Solving for the quantum state of photons reflected from the cavity}

%%%%%%%%%%%%%%%%%%%%%%%%%%%%%%%%%%%%%%%%%%%%%%%
%%%%%%%%%%%%%%%%%%%%%%%%%%%%%%%%%%%%%%%%%%%%%%%

Here we provide the details of solving Eqs.~(\ref{eqn30})-(\ref{eqn32}) for the coupled amplitudes of the external quantized field, quantized cavity field, and the states of QEs. After the substitution
$C_{1k} = s_{k} e^{- i\omega_{k}t}$,
$C_{c} = c e^{- i\omega_{c}t}$,
$C_{e} = f e^{- i\frac{W_{e}}{\hbar}t}$,
$- \frac{i}{\hbar}\mathfrak{R}_{e}\left( t \right)\mathbf{=}r_{e} e^{- i\omega_{c}t}$;
we obtain
\begin{eqnarray}
\frac{ds_{k}}{dt} = i\mathfrak{L}^{*}c e^{i\Delta_{k}t}\label{eqn47}\\
\frac{dc}{dt} + \frac{\mu_{c}}{2}c = i\sum_{k}{\mathfrak{L}s_{k}  e^{- i\Delta_{k}t}} + i\Omega_{c}^{*}f  e^{- i\Delta_{e}t}\label{eqn48}\\
\frac{df}{dt} + \frac{\gamma}{2}f = i\Omega_{c}c e^{i\Delta_{e}t} + r_{e}\left( t \right) e^{i\Delta_{e}t}\label{eqn49},
\end{eqnarray}
where $\Delta_{k} = \omega_{k} - \omega_{c}$, $\Delta_{e} = \frac{W_{e}}{\hbar} - \omega_{c}$. From (\ref{eqn47}) and (\ref{eqn49}) we get
\begin{eqnarray}
s_{k} = s_{k}\left( 0 \right) + i\mathfrak{L}^{*}\int_{0}^{t}{c{\left( \tau \right)e}^{+ i\Delta_{k}\tau}d\tau},\label{eqn50}\\
f = i\Omega_{c}e^{-\frac{\gamma}{2}t}\int_{0}^{t}{c\left( \tau \right)e^{\left( i\Delta_{e} + \frac{\gamma}{2} \right)\tau}}d\tau\nonumber\\+ e^{-\frac{\gamma}{2}t}\int_{0}^{t}{r_{e}\left( \tau \right)e^{\left( i\Delta_{e} + \frac{\gamma}{2} \right)\tau}}\label{eqn51},
\end{eqnarray}
where we assumed $f\left( 0 \right) = C_{e}\left( 0 \right) = 0$. Substituting Eqs.~(\ref{eqn50}) and (\ref{eqn51}) into Eq.~(\ref{eqn48}) yields
\begin{eqnarray}
\frac{dc}{dt} + \frac{\mu_{c}}{2}c = i\mathfrak{L}\sum_{k}{s_{k}\left( 0 \right)e^{- i\Delta_{k}t}}\nonumber\\- \left|\mathfrak{L} \right|^{2}\sum_{k}\int_{0}^{t}{c\left( \tau \right)e^{- i\Delta_{k}\left( t - \tau \right)}d\tau} -\nonumber\\
 {\left|\Omega_{c} \right|}^{2}\int_{0}^{t}{e^{- \left( i\Delta_{e} + \frac{\gamma}{2} \right)\left( t - \tau \right)}c\left( \tau \right)}d\tau\nonumber \\+ i\Omega_{c}^{*}\int_{0}^{t}e^{- \left( i\Delta_{e} + \frac{\gamma}{2} \right)\left( t - \tau \right)}r_{e}\left( \tau \right)d\tau
 \label{eqn52}
\end{eqnarray}
The last three terms on the rhs of Eq.~(\ref{eqn52}) are proportional to the convolutions
\begin{eqnarray}
\int_{0}^{t}{e^{- \left( i\Delta_{e} + \frac{\gamma}{2} \right)\left( t - \tau \right)}c\left( \tau \right)}d\tau = R\left( t \right)\otimes c\left( t \right),\nonumber\\
\int_{0}^{t}e^{- \left( i\Delta_{e} + \frac{\gamma}{2} \right)\left( t - \tau \right)}r_{e}\left( \tau \right)d\tau = R\left( t \right)\otimes r_{e}\left( t \right),\nonumber\\
\sum_{k}\int_{0}^{t}{c\left( \tau \right)e^{- i\Delta_{k}\left( t - \tau \right)}d\tau} = D\left( t \right)\otimes c\left( t \right),\nonumber
\end{eqnarray}
where we defined
$A\left( t \right)\otimes B\left( t \right) = \int_{0}^{t}{A\left( t - \tau \right)B\left( \tau \right)d\tau} = \int_{0}^{t}{A\left( \tau \right)B\left( t - \tau \right)d\tau}$,
$R\left( t \right) = e^{- \left( i\Delta_{e} + \frac{\gamma}{2} \right)t}$,
$D\left( t \right) = \sum_{k}e^{- i\Delta_{k}t}$.
Now we use the Laplace transform
\begin{equation}
c_{p} = \int_{0}^{\infty}{dt}e^{- pt}c\left( t \right),
c\left( t \right) = \frac{1}{2\pi i}\int_{x - i\infty}^{x + i\infty}{c_{p}e^{pt}dp}\nonumber
\end{equation}
where it is sufficient to consider $x>0$. Using the Laplace
transform of the convolutions, Eq.~(\ref{eqn52}) yields
\begin{equation}
pc_{p} - c\left( 0 \right) + \frac{\mu_{c}}{2}c_{p} + \left| \Omega_{c} \right|^{2}R_{p}c_{p} + \left| \mathfrak{L} \right|^{2}D_{p}c_{p} = i\mathfrak{L}F_{p} + i\Omega_{c}^{*}R_{p}r_{p}\nonumber
\end{equation}
where 
\begin{eqnarray}
R_{p} = \int_{0}^{\infty}e^{- \left( i\Delta_{e} + \frac{\gamma}{2} + p \right)t}dt = \frac{1}{i\Delta_{e} + \ \frac{\gamma}{2}\  + p},\nonumber\\
D_{p} = \int_{0}^{\infty}{\sum_{k}^{}e^{\left( - i\Delta_{k} - p \right)t}dt} = \sum_{k}\frac{1}{i\Delta_{k} + p},\nonumber\\
F_{p} = \int_{0}^{\infty}{\sum_{k}^{}{s_{k}\left( 0 \right)e^{\left( - i\Delta_{k} - p \right)t}dt}} = \sum_{k}\frac{s_{k}\left( 0 \right)}{i\Delta_{k} + p},\nonumber\\
r_{p} = \int_{0}^{\infty}{r_{e}\left( t \right)e^{- pt}dt}.\nonumber
\end{eqnarray}
Taking into account the initial condition $c\left( 0 \right) = C_{c}\left( 0 \right) = 0$, we get
\begin{eqnarray}
c_{p} &=& i\mathfrak{L}\left( \sum_{k}\frac{s_{k}\left( 0 \right)\left( i\Delta_{k} + p \right)^{-1}}{p+\frac{\mu_{c}}{2} +  \frac{\left| \Omega_{c} \right|^{2}}{i\Delta_{e} + \frac{\gamma}{2} + p} + \sum_{q}\frac{\left| \mathfrak{L} \right|^{2}}{i\Delta_{q} + p} } \right)\nonumber\\&+& i\Omega_{c}^{*}\frac{r_{p}\left( i\Delta_{e} + \frac{\gamma}{2} + p \right)^{-1}}{ p  + \frac{\mu_{c}}{2} + \frac{\left| \Omega_{c} \right|^{2}}{i\Delta_{e} + \frac{\gamma}{2} + p} + \sum_{q}\frac{\left| \mathfrak{L} \right|^{2}}{i\Delta_{q} + p}}.
\label{eqn53}
\end{eqnarray}
In Eq.~(\ref{eqn53}) we can go from summation $\sum_{q}\frac{\left|\mathfrak{L}\right|^{2}}{i\Delta_{q} + p}$
over wavenumbers to integration over frequencies,
\begin{equation}
\sum_{q}\frac{\left|\mathfrak{L} \right|^{2}}{i\Delta_{q} + p}\Longrightarrow\frac{L}{2\pi\upsilon_{g}}\int_{-\infty}^{\infty}{\frac{\left|\mathfrak{L}\right|^{2}}{i\Delta_{q} + p}d\Delta_{q}},
\label{eqn54}
\end{equation}
where, as usual, the analytic continuation of the complex \emph{p} plane
to the region $\text{Re}\left\lbrack p \right\rbrack \leq 0$
corresponds to the counter-clockwise integration path around the poles
in the integrals: $\int_{- \infty}^{\infty}{\frac{d\Delta_{q}}{i\Delta_{q} + p} = \pi}$. As a result we obtain
\begin{eqnarray}
c_{p} = i\mathfrak{L}\left(\sum_{k}\frac{s_{k}\left(0\right)\left(p_{e} + p\right)}{\left(i\Delta_{k} + p \right)\left\lbrack \left(p + \kappa_{\Sigma}\right)\left(p_{e} + p\right) + \left|\Omega_{c}\right|^{2}\right\rbrack}\right)\nonumber\\ + i\Omega_{c}^{*}\frac{r_{p}}{\left(p + \kappa_{\Sigma}\right)\left( p_{e} + p\right) + \left|\Omega_{c}\right|^{2}}\nonumber\\
\label{eqn55}
\end{eqnarray}
%\begin{widetext}
%\begin{eqnarray}
%c_{p} = &i\mathfrak{L}\left(\sum_{k}\frac{s_{k}\left(0\right)\left(p_{e} + p\right)}{\left(i\Delta_{k} + p\right)\left\lbrack \left(p + \kappa_{\Sigma}\right)\left(p_{e} + p\right) + \left|\Omega_{c}\right|^{2}\right\rbrack}\right) + i\Omega_{c}^{*}\frac{r_{p}}{\left(p + \kappa_{\Sigma}\right)\left( p_{e} + p\right) + \left|\Omega_{c}\right|^{2}}
%\label{eqn55}
%\end{eqnarray}
%\end{widetext}
where
\begin{equation}
p_{e} = \frac{\gamma}{2} + i\Delta_{e},
\kappa_{\Sigma} = \frac{\mu_{c}}{2} + \kappa,
\kappa = \frac{L\left| \mathfrak{L} \right|^{2}}{2\upsilon_{g}}.
\label{eqn56}
\end{equation}

Note that the value of $\mathfrak{L}$ should be proportional to the
normalization field of the quantized mode of the ray bundle (\ref{eqn25}), from
where we can parameterize it as
$\left| \mathfrak{L} \right| = \frac{c}{\sqrt{L{l_{d}}}}$, where the
quantity $l_{d}$ has the dimension of length and can be obtained from
solving the classical electrodynamics problem of cavity excitation by an
external classical field. For example, for the Fabry-Perot cavity
excited through a semi-transparent mirror with amplitude transmission
coefficient $T$ we obtain
$l_{d}\sim\left|{T} \right|^{2}l_{c}$, where $l_{c}$
is the distance between the mirrors.
The result is
\begin{equation}
\kappa = \frac{c^{2}}{2\upsilon_{g}l_{d}}, 
\label{eqn57}
\end{equation}
where the parameter $\kappa$ does not depend on the arbitrary
quantization volume as it should be. The value of $\kappa$ describes
the part of the cavity decay rate which is due to outcoupling into the
modes of the ray bundle sketched in Fig.~1, whereas the quantity
$\kappa_{\Sigma}$ is the total decay rate of the cavity mode.
To calculate the inverse Laplace transform based on Eq.~(\ref{eqn55})
it is convenient to transform it as
\begin{eqnarray}
c_{p} &=&i \mathfrak{L} \sum_{k}{ s_{k}\left( 0 \right)}\times
\Biggr[\frac{(p_{e} - i\Delta_{k})(p+i\Delta_{k})^{-1}}{\left( P_{1} + i\Delta_{k} \right)\left( i\Delta_{k} + P_{2} \right)}\nonumber\\& +&\frac{(P_{1} + p_{e})(p - P_{1})^{-1}}{\left( P_{1} + i\Delta_{k} \right)\left( P_{1} - P_{2} \right)} - \frac{(P_{2} + p_{e})(p - P_{2})^{-1}}{\left( P_{2} + i\Delta_{k} \right)\left( P_{1} - P_{2} \right)}\Biggr]\nonumber\\
&+&\frac{i\Omega_{c}^{*}r_{p}}{P_{1} - P_{2}}\left( \frac{1}{p - P_{1}} - \frac{1}{p - P_{2}} \right)
\label{eqn58}
\end{eqnarray}
%\begin{widetext}
%\begin{eqnarray}
%c_{p} = i\sum_{k}{\mathfrak{L}_{k}s_{k}\left( 0 \right)} \times
%\left\lbrack \frac{(p_{e} - i\Delta_{k})(p+i\Delta_{k})^{-1}}{\left( P_{1} + i\Delta_{k} \right)\left( i\Delta_{k} + P_{2} \right)} + \frac{(P_{1} + p_{e})(p - P_{1})^{-1}}{\left( P_{1} + i\Delta_{k} \right)\left( P_{1} - P_{2} \right)} - \frac{(P_{2} + p_{e})(p - P_{2})^{-1}}{\left( P_{2} + i\Delta_{k} \right)\left( P_{1} - P_{2} \right)}\right\rbrack\nonumber \\
%+\frac{i\Omega_{c}^{*}r_{p}}{P_{1} - P_{2}}\left( \frac{1}{p - P_{1}} - \frac{1}{p - P_{2}} \right)
%\label{eqn58}
%\end{eqnarray}
%\end{widetext}
where
\begin{equation}
P_{1,2} = - \frac{\kappa_{\Sigma} + p_{e}}{2} \pm \sqrt{\left( \frac{\kappa_{\Sigma} - p_{e}}{2} \right)^{2} - \left| \Omega_{c} \right|^{2}}
\label{eqn59}
\end{equation}
are the roots of equation $\left(p +\kappa_{\Sigma}\right)\left(p_{e} + p\right)+\left|\Omega_{c} \right|^{2} = 0$.
Applying the inverse Laplace transform to Eq.~(\ref{eqn58}), we get
\begin{eqnarray}
c &=& i\mathfrak{L}\sum_{k}s_{k}\left( 0 \right)\Biggr[\frac{\left( p_{e} - i\Delta_{k} \right)e^{- {i\Delta}_{k}t}}{\left( P_{1} + i\Delta_{k} \right)\left( P_{2} + i\Delta_{k} \right)}\nonumber\\& +& \frac{\left( P_{1} + p_{e} \right)e^{P_{1}t}}{\left( P_{1} + i\Delta_{k} \right)\left( P_{1} - P_{2} \right)} - \frac{\left( P_{2} + p_{e} \right)e^{P_{2}t}}{\left( P_{2} + i\Delta_{k} \right)\left( P_{1} - P_{2} \right)} \Biggr]\nonumber\\& + &\delta c\left( t \right)
\label{eqn60}
\end{eqnarray}
%\begin{widetext}
%\begin{equation}
%c = i\mathfrak{L}\sum_{k}{s_{k}\left( 0 \right)\left\lbrack \frac{\left( p_{e} - i\Delta_{k} \right)e^{- {i\Delta}_{k}t}}{\left( P_{1} + i\Delta_{k} \right)\left( P_{2} + i\Delta_{k} \right)} + \frac{\left( P_{1} + p_{e} \right)e^{P_{1}t}}{\left( P_{1} + i\Delta_{k} \right)\left( P_{1} - P_{2} \right)} - \frac{\left( P_{2} + p_{e} \right)e^{P_{2}t}}{\left( P_{2} + i\Delta_{k} \right)\left( P_{1} - P_{2} \right)} \right\rbrack} + \delta c\left( t \right)
%\label{eqn60}
%\end{equation}
%\end{widetext}
where
\begin{equation}
\delta c\left( t \right) = \frac{i\Omega_{c}^{*}}{P_{1} - P_{2}}r_{e}\left( t \right)\otimes\left( e^{P_{1}t} - e^{P_{2}t} \right).
\label{eqn61}
\end{equation}

Next we consider a narrowband signal, when
\begin{equation}
\left| \Delta_{k} - \Delta_{0} \right| \leq \mathrm{\Delta}\omega \ll \left| P_{1,2} \right|.
\label{eqn62}
\end{equation}
In this case one can put $\Delta_{k} = \Delta_{0}$ in Eq.~(\ref{eqn60})
everywhere except the exponential factor in the first term in the
brackets. When performing the summation
$\sum_{k}{s_{k}\left( 0 \right)\left\lbrack \ldots \right\rbrack}$
we can neglect the last two terms in the brackets in Eq.~(\ref{eqn60}) because
after substituting \(\Delta_{k} = \Delta_{0}\) they do not depend on
index \emph{k} and therefore, due to Eq.~(\ref{eqn46}) only the first term in the
brackets contributes. As a result,
\begin{equation}
c = i\mathfrak{L}\sum_{k}{s_{k}\left( 0 \right)\frac{\left( p_{e} - i\Delta_{0} \right)e^{- {i\Delta}_{k}t}}{\left( P_{1} + i\Delta_{0} \right)\left( P_{2} + i\Delta_{0} \right)}} + {\delta}c\left( t \right). 
\label{eqn63}
\end{equation}

Let us specify the conditions for going from Eq.~(\ref{eqn60}) to Eq.~(\ref{eqn63}). In the
strong coupling regime, when
$\left| \Omega_{c} \right| \gg \kappa_{\Sigma},\left| p_{e} \right|$,
we obtain $P_{1,2} \approx \pm i\left| \Omega_{c} \right|$; in this
case the inequality (\ref{eqn62}) is valid if $\mathrm{\Delta}\omega \ll \left| \Omega_{c} \right|$. In the weak
coupling regime, when
$\left|\Omega_{c} \right| \ll \kappa_{\Sigma},\left| p_{e} \right|$,
one of the roots in Eq.~(\ref{eqn59}) approaches $-p_e$. Let us take $P_{1} \approx {- p}_{e}$. In this case the expression in
brackets in Eq.~(\ref{eqn60}) is reduced to the form independent on the root
$P_{1}$:
\begin{eqnarray}
\frac{\left( p_{e} - i\Delta_{k} \right)e^{- {i\Delta}_{k}t}}{\left( P_{1} + i\Delta_{k} \right)\left( P_{2} + i\Delta_{k} \right)} + \frac{\left( P_{1} + p_{e} \right)e^{P_{1}t}}{\left( P_{1} + i\Delta_{k} \right)\left( P_{1} - P_{2} \right)}\nonumber\\ - \frac{\left( P_{2} + p_{e} \right)e^{P_{2}t}}{\left( P_{2} + i\Delta_{k} \right)\left( P_{1} - P_{2} \right)} \approx - \frac{e^{- {i\Delta}_{k}t} - e^{P_{2}t}}{P_{2} + i\Delta_{k}},\nonumber
\end{eqnarray}
i.e., in this case the condition (\ref{eqn62}) is modified (it does not depend on $P_{1}$):
\begin{equation}
\left| \Delta_{k} - \Delta_{0} \right| \leq \mathrm{\Delta}\omega \ll \left| P_{2} \right| \approx \kappa_{\Sigma}.
\label{eqn64}
\end{equation}
The transition from Eq.~(\ref{eqn60}) to Eq.~(\ref{eqn63}) is possible, in particular, in the range of parameters
$\gamma \ll \mathrm{\Delta}\omega \ll \max\left\lbrack \left| \Omega_{c} \right|,\kappa_{\Sigma} \right\rbrack$,
i.e., when
$\gamma \ll \max\left\lbrack \left| \Omega_{c} \right|,\kappa_{\Sigma} \right\rbrack$.
The pulse duration $\sim{\mathrm{\Delta}\omega}^{- 1}$ can be much
shorter than the relaxation time of QEs $\sim\gamma^{- 1}$.

Next we substitute Eq.~(\ref{eqn63}) into Eq.~(\ref{eqn50}) and again assume a
relatively narrow bandwidth of the signal $\mathrm{\Delta}\omega$.
Returning to the original variables $C_{1k} = s_{k} \cdot e^{- i\omega_{k}t}$, we obtain
\begin{widetext}
\begin{equation}
C_{1k}\left( t \right) \approx \left( C_{1k}\left( 0 \right) - \frac{\left( p_{e} - i\Delta_{0} \right)\left| \mathfrak{L} \right|^{2}}{\left( P_{1} + i\Delta_{0} \right)\left( P_{2} + i\Delta_{0} \right)}\int_{0}^{t}{\sum_{q}{C_{1q}\left( 0 \right)e^{i\left( \omega_{k} - \omega_{q} \right)\tau}d\tau}} \right)e^{- i\omega_{k}t} + \delta C_{1k}\left( t \right),
\label{eqn65}
\end{equation}
%\end{widetext}
where
\begin{eqnarray}
\delta C_{1k}\left( t \right) = \frac{i\mathfrak{L}^{*}\Omega_{c}^{*}}{\hbar\left( P_{1} - P_{2} \right)}\times\left\{ \left\{\mathfrak{R}_{e}\left( t \right)\otimes\left\lbrack e^{- i\omega_{c}t}\left( e^{P_{1}t} - e^{P_{2}t} \right) \right\rbrack \right\}\otimes e^{- i\omega_{k}t} \right\}
\label{eqn66}
\end{eqnarray}
%\begin{eqnarray}
%\delta C_{1k}\left( t \right) = \frac{i\mathfrak{L}^{*}\Omega_{c}^{*}}{\hbar\left( P_{1} - P_{2} \right)}\times\nonumber\\\left\{ \left\{\mathfrak{R}_{e}\left( t \right)\otimes\left\lbrack e^{- i\omega_{c}t}\left( e^{P_{1}t} - e^{P_{2}t} \right) \right\rbrack \right\}\otimes e^{- i\omega_{k}t} \right\}
%\label{eqn66}
%\end{eqnarray}
Using Eq.~(\ref{eqn45}), when $t > T_{1}$ in Eq.~(\ref{eqn65}), in the last term in parentheses on the rhs we can extend the integration to $\pm\infty$.
This gives
%\begin{widetext}
\begin{equation}
C_{1k}\left( t > T_{1} \right) \approx \left( C_{1k}\left( 0 \right) - \frac{\left( p_{e} - i\Delta_{0} \right)\left| \mathfrak{L} \right|^{2}}{\left( P_{1} + i\Delta_{0} \right)\left( P_{2} + i\Delta_{0} \right)}\sum_{q}{C_{1q}\left( 0 \right)\int_{- \infty}^{\infty}{e^{i\left( \omega_{k} - \omega_{q} \right)\tau}d\tau}} \right)e^{- i\omega_{k}t} + \delta C_{1k}\left( t \right).\nonumber
\end{equation}
%\end{widetext}
Again going from summation to integration over frequencies,
$\sum_{q}{\left( \ldots \right) \Longrightarrow \frac{L}{2\pi\upsilon_{g}}\int_{- \infty}^{\infty}{\left( \ldots \right)d\omega_{q}}}$, we arrive at
\begin{equation}
C_{1k}\left( t > T_{1} \right) = C_{1k}\left( 0 \right)\left( 1 - \frac{2\left( p_{e} - i\Delta_{0} \right)}{\left( P_{1} + i\Delta_{0} \right)\left( P_{2} + i\Delta_{0} \right)}\right)e^{- i\omega_{k}t} + \delta C_{1k}\left( t \right).
\label{eqn67}
\end{equation}
\end{widetext}
The condition $t > T_{1}$ corresponds to the time when the external
pulse has already propagated away from the cavity upon reflection; at
that time the cavity field and any excitation of the QEs has already
dissipated. 

The fluctuation term $\delta C_{1k}\left( t \right)$ can be usually neglected when the temperature is much lower than the optical transition frequency in QEs. We neglected it in the analysis in Sec.~IV. The effect of fluctuations can become important in the presence of of strong dephasing processes. Their impact is analyzed separately in Appendix C below.

%%%%%%%%%%%%%%%%%%%%%

%%%%%%%%%%%%%%%%%%%%%%%%%%%%%%%%%%%%%%%%%%%%%%%%%%%%%%

%%%%%%%%%%%%%%%%%%%%%%%%%%%%%%%%%%%%%%%%%%%
%%%%%%%%%%%%%%%%%%%%%%%%%%%%%%%%%%%%%%%%%%%%
\section{Fluctuations due to pure dephasing}
%%%%%%%%%%%%%%%%%%%%%%%%%%%%%%%%%%%%%%%%%%%%
%%%%%%%%%%%%%%%%%%%%%%%%%%%%%%%%%%%%%%%%%%%%

The effect of pure dephasing needs a separate treatment because in this case the
Langevin noise term is present directly in the equation of motion for
the excited states even at zero temperature of the reservoir (see the
second of Eq.~(\ref{eqn36}). To estimate its contribution we will use
Eq.~(\ref{eqn66}) and (\ref{eqn42}). At exact resonance between the cavity field
and the QEs, when $\Delta_{e} = \frac{W_{e}}{\hbar} - \omega_{c} = 0$, and in the
strong coupling regime
$\left| \Omega_{c} \right| \gg \kappa_{\Sigma},\gamma$ we obtain
%\begin{equation}
%\left|\delta C_{1k} \right|^{2} \approx 2\gamma^{\left(el\right)}\left| \mathfrak{L} \right|^{2} \times
%\times\int_{0}^{t}d\tau\int_{0}^{t}{d\tau^{\prime}}e^{- i\mathrm{\Delta}_{k}\left( \tau - \tau^{\prime} \right)}\int_{0}^{\tau}d\chi\overline{\left| C_{e}\left( \tau - \chi \right) \right|^{2}}e^{- \Gamma\left\lbrack 2\chi - \left( \tau - \tau^{\prime} \right) \right\rbrack}\sin\left( \left| \Omega_{c} \right|\chi \right)\sin\left\{ \left| \Omega_{c} \right|\left\lbrack \chi - \left( \tau - \tau^{\prime} \right) \right\rbrack \right\}
%\label{eqnb1}
%\end{equation}
\begin{eqnarray}
\left|\delta C_{1k} \right|^{2} \approx 2\gamma^{\left(el\right)}\left| \mathfrak{L} \right|^{2}
\int_{0}^{t}d\tau\int_{0}^{t}{d\tau^{\prime}}e^{- i\mathrm{\Delta}_{k}\left( \tau - \tau^{\prime} \right)}\times\nonumber\\
\int_{0}^{\tau}d\chi\overline{\left| C_{e}\left( \tau - \chi \right) \right|^{2}}e^{- \Gamma\left\lbrack 2\chi - \left( \tau - \tau^{\prime} \right) \right\rbrack}\sin\left( \left| \Omega_{c} \right|\chi \right)\times\nonumber\\
\sin\left\{ \left| \Omega_{c} \right|\left\lbrack \chi - \left( \tau - \tau^{\prime} \right) \right\rbrack \right\}
\label{eqnb1}
\end{eqnarray}
%\begin{eqnarray}
%\left|\delta C_{1k} \right|^{2} \approx 2\gamma^{\left(el\right)}\left| \mathfrak{L} \right|^{2} \times
%\int_{0}^{t}d\tau\int_{0}^{t}{d\tau^{\prime}}e^{- i\mathrm{\Delta}_{k}\left( \tau - \tau^{\prime} \right)}\int_{0}^{\tau}d\chi\overline{\left| C_{e}\left( \tau - \chi \right) \right|^{2}}e^{- \Gamma\left\lbrack 2\chi - \left( \tau - \tau^{\prime} \right) \right\rbrack}\sin\left( \left| \Omega_{c} \right|\chi \right)\sin\left\{ \left| \Omega_{c} \right|\left\lbrack \chi - \left( \tau - \tau^{\prime} \right) \right\rbrack \right\}
%\label{eqnb1}
%\end{eqnarray}
where
$\Gamma = \frac{1}{2}\left( \kappa_{\Sigma} + \frac{\gamma}{2} \right) = \frac{1}{2}\left( \kappa\  + \ \frac{\mu_{c}}{2} + \frac{\gamma}{2} \right)$.
Now we sum the noise term over the modes of the ray bundle, using
$\sum_{k}\left( \ldots \right) \Longrightarrow \frac{L}{2\pi\upsilon_{g}}\int_{- \infty}^{\infty}{\left( \ldots \right)d\mathrm{\Delta}_{k}}$ and
$\int_{- \infty}^{\infty}{e^{- i\mathrm{\Delta}_{k}\left( \tau - \tau^{\prime} \right)}d\mathrm{\Delta}_{k}} = 2\pi\delta\left( \tau - \tau^{\prime} \right)$, arriving at
\begin{eqnarray}
\sum_{k}\left| \delta C_{1k} \right|^{2} \approx 4\gamma^{\left(el\right)}\kappa\int_{0}^{t}d\tau\int_{0}^{\tau}d\chi\overline{\left|C_{e}\left(\tau - \chi \right)\right|^{2}}\nonumber\\\times e^{- 2\Gamma\chi}\frac{1 - \cos\left( 2\left| \Omega_{c} \right|\chi \right)}{2}
\label{eqnb2}
\end{eqnarray}
Next, we take into account that in the strong coupling regime
$\left|\Omega_{c}\right| \gg \Gamma$, so that we can neglect
a rapidly oscillating term
$\propto \cos\left( 2\left| \Omega_{c} \right|\chi \right)$ in
Eq.~(\ref{eqnb2}). We also take into account that the time duration of the
signal, $\left|C_{e}\left(t\right)\right|^{2}$, corresponds to the
inverse bandwidth $\sim\frac{1}{\mathrm{\Delta}\omega}$, where we
assume $\Delta \omega \ll \Gamma$. The result is
\begin{equation}
\sum_{k}\left|\delta C_{1k} \right|^{2} \approx \frac{\gamma^{\left(el\right)}\kappa}{\Gamma}\int_{0}^{t}{d\tau}\overline{\left| C_{e}\left( \tau \right) \right|^{2}}
\label{eqnb3}
\end{equation}

Now we estimate the value of
$\int_{0}^{t}d\tau\overline{\left| C_{e}\left( \tau \right) \right|^{2}}$
using Eqs.~(\ref{eqn31}) and (\ref{eqn32}). Within the perturbation theory, we can
assume there that $C_{1k} = C_{1k}\left( 0 \right)e^{- i\omega_{k}t}$
and $\mathfrak{R}_{e}\left( t \right) \rightarrow 0$. We
furthermore include the radiation of the cavity field into the modes of
the ray bundle by replacing
$\frac{\mu_{c}}{2}\Rightarrow\kappa_{\Sigma} = \kappa + \frac{\mu_{c}}{2}$.
This yields
%\begin{equation}
%C_{e}\left( t \right)e^{i\omega_{c}t} = \sum_{k}\frac{\mathfrak{L}\Omega_{c}C_{1k}\left( 0 \right)e^{- i\mathrm{\Delta}_{k}t}}{\left( \Delta_{k} + i\Gamma - \sqrt{\left| \Omega_{c} \right|^{2} - \frac{1}{4}\left( \kappa_{\Sigma} - \frac{\gamma}{2} \right)^{2}} \right)\left( \Delta_{k} + i\Gamma + \sqrt{\left| \Omega_{c} \right|^{2} - \frac{1}{4}\left( \kappa_{\Sigma}  - \frac{\gamma}{2} \right)^{2}}\right)},
%\label{eqnb4}
%\end{equation}
\begin{equation}
C_{e}\left( t \right)e^{i\omega_{c}t} = \sum_{k}\frac{\mathfrak{L}\Omega_{c}C_{1k}\left( 0 \right)e^{- i\mathrm{\Delta}_{k}t}}{\left( \Delta_{k} + i\Gamma - P^{\prime}\right)\left( \Delta_{k} + i\Gamma + P^{\prime}\right)},
\label{eqnb4} 
\end{equation}
where, $P^{\prime}=\sqrt{\left|\Omega_{c}\right|^{2} - \frac{1}{4}\left(\kappa_{\Sigma} - \frac{\gamma}{2} \right)^{2}}$. In the strong coupling regime under the condition
$\mathrm{\Delta}\omega \ll \left| \Omega_{c} \right|$ we obtain
\begin{equation}
\left| C_{e}\left( t \right) \right|^{2} \approx \frac{2\kappa\left|\Omega_{c} \right|^{2}\sum_{k}\left| C_{1k}\left( 0 \right) \right|^{2}}{\left| \left(\Delta_{0} + i\Gamma - \left| \Omega_{c} \right| \right)\left(\Delta_{0} + i\Gamma + \left| \Omega_{c} \right| \right)\right|^{2}}.
\label{eqnb5}
\end{equation}
Substituting Eq.~(\ref{eqnb5}) into Eq.~(\ref{eqnb3}) and taking into account that
$\left| C_{e}\left( t \right) \right|^{2} \neq 0$ only during
$T_{0} < t < T_{1}$, we have
\begin{equation}
\left|\delta C_{1k}\left(t > T_{1} \right)\right|^{2}\approx \frac{2\gamma^{\left(el\right)}\kappa^{2}\left| \Omega_{c} \right|^{2}\Gamma^{-1}\sum_{k}\left| C_{1k}\left( 0 \right) \right|^{2}}{\left|\left(\Delta_{0} + i\Gamma - \left| \Omega_{c}\right| \right)\left(\Delta_{0} + i\Gamma + \left| \Omega_{c} \right| \right)\right|^{2}}.
\label{eqnb6}
\end{equation}
For $\Delta_{0} = 0$ this becomes
\begin{equation}
\sum_{k}\left| \delta C_{1k}\left( t > T_{1} \right) \right|^{2} \approx \frac{4\gamma^{\left(el\right)}\kappa^{2}}{\left(\kappa + \frac{\mu_{c}}{2}\  + \frac{\gamma}{2} \right)\left| \Omega_{c} \right|^{2}}\sum_{k}^{}\left| C_{1k}\left( 0 \right) \right|^{2}.
\label{eqnb7}
\end{equation}
At the same time, in Sec.~IV we obtained that the main
component of the signal in the strong coupling regime satisfies
$C_{1k}\left( t > T_{1} \right) \approx C_{1k}\left( 0 \right)e^{- i\omega_{k}t}$.
A relatively small contribution of the fluctuations due to pure
dephasing can be explained by the fact that the central frequency of the
external pulse, which is resonant to the frequency of an empty cavity,
turned out to be strongly detuned from resonant frequencies of the
strongly coupled system. If, however, we choose the carrier frequency in resonance with one
of the coupling-shifted frequencies
$\omega_{0} \approx \omega_{c} \pm \left| \Omega_{c} \right|$, the
fluctuations become much stronger:
\begin{equation}
\sum_{k}\left| \delta C_{1k}\left( t > T_{1} \right) \right|^{2} \approx \frac{4\gamma^{\left( \text{el} \right)}\kappa^{2}}{\left( \kappa\  + \ \frac{\mu_{c}}{2}\  + \ \frac{\gamma}{2} \right)^{3}}\sum_{k}^{}\left| C_{1k}\left( 0 \right) \right|^{2}.
\label{eqnb8}
\end{equation}
In particular, in the critical coupling regime 
we obtain
\begin{equation}
\sum_{k}\left| \delta C_{1k}\left(t > T_{1}\right) \right|^{2} \approx \frac{\gamma^{\left(el\right)}}{\mu_{c} + \gamma_{e} + 2\gamma^{\left(el\right)}}\sum_{k}\left| C_{1k}\left(0\right)\right|^{2}
\label{eqnb9}
\end{equation}
In this case the noise due to pure dephasing is small only in the limit
of strong field losses, when
$\kappa_{\Sigma} \gg \gamma^{\left(el\right)}$.

%%%%%%%%%%%%%%%%%%%%%%%%%%%%%%%%%%%%%%%%%%%%%%%
\section{Quantum state of the field in different polarization
bases}

%%%%%%%%%%%%%%%%%%%%%%%%%%%%%%%%%%%%%%%%%%%%%%

Consider an operator of the electric field along the ray bundle,
\begin{equation}
\hat{\mathbf{E}}_{out}\left(\mathbf{\xi}_{\mathbf{\bot}},s \right)\mathbf=\left(\mathbf{e}_{1}\hat{c}_{e1} + \mathbf{e}_{2}{\hat{c}}_{e2}\right)E_{0}\left(\mathbf{\xi}_{\mathbf{\bot}}\right)e^{iks} + h.c.
\label{eqnc1}
\end{equation}
Here \emph{s} is the coordinate along the \textquotedblleft{central ray}\textquotedblright of the ray
bundle, $\mathbf{e}_{1,2}$ are orthogonal vectors of polarizations of
the two normal modes in the plane normal to the ray,
$\mathbf{\xi}_\mathbf{\bot}$ and
$\left|E_{0}\left(\mathbf{\xi}_{\mathbf{\bot}},s\right)\right|^{2}$
are the coordinates and distribution of intensity in this
plane, ${\hat{c}}_{e_{1,2}}$ are annihilation operators of the Fock
states corresponding to these modes. We assume polarization degeneracy
which permits an arbitrary choice of the basis $\mathbf{e}_{1,2}$.
Consider transition to a new basis $\mathbf{g}_{1,2}$:
\begin{equation}
\begin{pmatrix}
\mathbf{g}_{1} \\
\mathbf{g}_{2} \\
\end{pmatrix} = \overset{\leftrightarrow}{t}\times 
\begin{pmatrix}
\mathbf{e}_{1}\\
\mathbf{e}_{2}\\
\end{pmatrix}
\label{eqnc2}
\end{equation}
in which the normalization is preserved,
$\left|\mathbf{e}_{1,2}\right|^{2}=\left|\mathbf{g}_{1,2} \right|^{2}=1$
and basis vectors remain orthogonal:
$\mathbf{e}_{1}\mathbf{e}_{2}^{\mathbf{\ast}} = \mathbf{g}_{1}\mathbf{g}_{2}^{\mathbf{\ast}} = 0$.
In this case the general form of the transformation matrix is
\begin{equation}
\overset{\leftrightarrow}{t}=
\begin{pmatrix}
t_{11} & t_{12}\\
t_{21} & t_{22}\\
\end{pmatrix} = 
\begin{pmatrix}
\alpha & \beta\\
-\beta^{*}e^{i\chi} & \alpha^{\ast}e^{i\chi}
\end{pmatrix}
\label{eqnc3}
\end{equation}
where $\left|\alpha\right|^{2} + \left|\beta\right|^{2} = 1$. 
Consider two examples relevant for this study.  

(a) Rotation of the Cartesian basis 
$\mathbf{e}_{1,2}$
by an angle $\phi$ around the central ray is given by
\begin{equation}
\overset{\leftrightarrow}{t}= \begin{pmatrix}
\cos\phi & \sin\phi \\
 - \sin\phi & \cos\phi \\
\end{pmatrix}, 
\label{eqnc4}
\end{equation}
i.e., $\alpha = \cos\phi$, $\beta = \sin\phi$, $e^{i\chi} = 1$.

(b) Transformation to the basis of circularly polarized modes
$\mathbf{g}_{1,2}=\frac{\mathbf{e}_{1}\mathbf{\pm}i\mathbf{e}_{2}}{\sqrt{2}}$,
when
\begin{equation}
\overset{\leftrightarrow}{t} = \frac{1}{\sqrt{2}}
\begin{pmatrix}
1 & i \\
1 & - i \\
\end{pmatrix}, 
\label{eqnc5}
\end{equation}
i.e., $\alpha = 1$, $\beta = i$, $e^{i\chi} = - i$.
How to transform the field operator (\ref{eqnc1}) in the polarization basis
$\mathbf{e}_{1,2}$ to the field operator in the polarization basis
$\mathbf{g}_{1,2}$? In the new basis 
\begin{equation}
\hat{\mathbf{E}}=\left( \mathbf{g}_{1}\hat{c}_{g_{1}} + \mathbf{g}_{2}\hat{c}_{g_2} \right)E_{0}\left(\mathbf{\xi}_{\mathbf{\bot}}\right)e^{iks} + h.c.,
\label{eqnc6}
\end{equation}
where from Eqs.~(\ref{eqnc1}), (\ref{eqnc2}) and (\ref{eqnc6}) we can get
\begin{eqnarray}
\hat{\mathbf{E}}=&\left\lbrack \mathbf{e_{1}}\left(t_{11}\hat{c}_{g_{1}}+t_{21}\hat{c}_{g_{2}}\right)+ \mathbf{e_{2}}\left(t_{12}\hat{c}_{g_{1}}+t_{22}\hat{c}_{g_{2}}\right)\right\rbrack\times\nonumber\\ &E_{0}(\mathbf{\xi_{\mathbf{\bot}}})e^{iks}+h.c.
\label{eqnc7}
\end{eqnarray}
Finally,
\begin{equation}
t_{11}{\hat{c}}_{g_{1}} + t_{21}{\hat{c}}_{g_{2}} = {\hat{c}}_{e_{1}},
t_{12}{\hat{c}}_{g_{1}} + t_{22}{\hat{c}}_{g_{2}} = {\hat{c}}_{e2}.
\label{eqnc8}
\end{equation}
It is easy to see that the transformation (\ref{eqnc3}) preserves all properties of the bosonic operators:
$\left\lbrack{\hat{c}}_{e_{1}},{\hat{c}}_{e_{1}}^{\dagger}\right\rbrack = \left\lbrack{\hat{c}}_{e_{2}},{\hat{c}}_{e_{2}}^{\dagger}\right\rbrack = \left\lbrack{\hat{c}}_{g_{1}},{\hat{c}}_{g_{1}}^{\dagger}\right\rbrack = \left\lbrack{\hat{c}}_{g_{2}},{\hat{c}}_{g_{2}}^{\dagger}\right\rbrack = \left|\alpha\right|^{2} + \left|\beta\right|^{2} = 1$
and $\left\lbrack{\hat{c}}_{e_{1}},{\hat{c}}_{e_{2}}^{\dagger}\right\rbrack = \left\lbrack {\hat{c}}_{g_{1}},{\hat{c}}_{g_{2}}^{\dagger}\right\rbrack = \alpha\beta^{\ast} - \beta^{\ast}e^{i\chi}\alpha e^{-i\chi} = 0$.
%\begin{eqnarray}
%\left\lbrack{\hat{c}}_{e_{1}},{\hat{c}}_{e_{1}}^{\dagger}\right\rbrack = \left\lbrack{\hat{c}}_{e_{2}},{\hat{c}}_{e_{2}}^{\dagger}\right\rbrack = \left\lbrack{\hat{c}}_{\gamma_{1}},%{\hat{c}}_{\gamma_{1}}^{\dagger}\right\rbrack = \nonumber\\\left\lbrack{\hat{c}}_{\gamma_{2}},
%{\hat{c}}_{\gamma_{2}}^{\dagger}\right\rbrack = \left|\alpha\right|^{2} + \left|\beta\right|^{2} = 1, \nonumber 
%\end{eqnarray}
%and
%\begin{equation}
%\left\lbrack{\hat{c}}_{e_{1}},{\hat{c}}_{e_{2}}^{\dagger}\right\rbrack = \left\lbrack{\hat{c}}_{\gamma %_{1}},{\hat{c}}_{\gamma_{2}}^{\dagger}\right\rbrack = \alpha\beta^{\ast} -
%\beta^{\ast}e^{i\chi}\alpha e^{-i\chi} = 0.\nonumber
%\end{equation}

Now consider the transformation of the state vector. We assume that in
the basis $\mathbf{e}_{1,2}$ the state vector was
\begin{eqnarray}
|\Psi^{\left(e\right)}\rangle = \sum_{p_{e_{1}},q_{e_{2}} = 0}^{\infty}{C_{p_{e_{1}}q_{e_{2}}}\left |p_{e_{1}}\right\rangle|q_{e_{2}}\rangle}\nonumber\\ = \sum_{p_{e_{1}},q_{e_{2}} = 0}^{\infty}{C_{p_{e_{1}}q_{e_{2}}}\frac{\left({\hat{c}}_{e_{1}}^{\dagger}\right)^{p_{e_{1}}}\left( {\hat{c}}_{e_{2}}^{\dagger}\right)^{q_{e_{2}}}}{\sqrt{p_{e_{1}}{!q}_{e_{2}}!}}\left|0_{\Sigma}\right\rangle},
\label{eqnc9}
\end{eqnarray}
where in all expressions of the type
$|p_{e_{1}}\rangle|q_{e_{2}}\rangle$ the
first term is for the states with polarization $\mathbf{e_{1}}$, and
the second place is for the states with polarization $\mathbf{e_{2}}$;
$\left|0_{\Sigma}\right\rangle$ is a vacuum state in any basis:
$|0_{\Sigma}\rangle = |0_{e_{1}}\rangle|0_{e_{2}}\rangle = |0_{g_{1}}\rangle|0_{g_{2}}\rangle$.
Substituting Eq.~(\ref{eqnc8}) into Eq.~(\ref{eqnc9}) we obtain the expression which allows
one to transfer the state vector into a different basis:
%\begin{equation}
%|\Psi^{\left( \gamma \right)}\rangle = \sum_{p_{e_{1}},q_{e_{1}} = 0}^{\infty}{C_{p_{e}q_{e}}\frac{\left( t_{11}^{*}{\widehat{c}}_{\gamma_{1}}^{\dagger} + t_{21}^{*}{\widehat{c}}_{\gamma_{2}}^{\dagger} \right)^{p_{e_{1}}}\left(t_{12}^{*}{\widehat{c}}_{\gamma_{1}}^{\dagger} + t_{22}^{*}{\widehat{c}}_{\gamma_{2}}^{\dagger} \right)^{q_{e_{2}}}}{\sqrt{p_{e_{1}}!q_{e_{1}}!}}\left|0_{\Sigma} \right\rangle}
%\label{eqnc10}
%\end{equation}
\begin{equation}
|\Psi^{\left(g\right)}\rangle = \sum_{p_{e_{1}},q_{e_{1}}=0}^{\infty}\frac{C_{p_{e}q_{e}}}{\sqrt{p_{e_{1}}!q_{e_{1}}!}}\frac{\left(t_{11}^{\ast}{\hat{c}}_{g_{1}}^{\dagger} + t_{21}^{\ast}{\hat{c}}_{g_{2}}^{\dagger}\right)^{p_{e_{1}}}}{\big(t_{12}^{\ast}{\hat{c}}_{g_{1}}^{\dagger} + t_{22}^{\ast}{\hat{c}}_{g_{2}}^{\dagger}\big)^{-q_{e_{1}}}}\left|0_{\Sigma}\right\rangle
\label{eqnc10}
\end{equation}
Therefore, each term
$|p_{e_{1}}\rangle|q_{e_{2}}\rangle$ is transformed as the following sum,
%\begin{widetext}
%\begin{eqnarray}
%|p_{e_{1}}\rangle|q_{e_{2}}\rangle \Longrightarrow \frac{1}{\sqrt{p_{e_{1}}!q_{e_{1}}!}}\sum_{k = %0}^{p_{e_{1}}}{\sum_{l = 0}^{q_{e_{2}}}\left\lbrack \sqrt{\left( p_{e_{1}} + q_{e_{2}} - k - l %\right)!\left( l + s \right)!}\begin{pmatrix}
%p_{e_{1}} \\
%k \\
%\end{pmatrix}\begin{pmatrix}
%q_{e_{2}} \\
%l \\
%\end{pmatrix} \times\right.\ }\nonumber\\
%\left. \ \left. \ {\left( t_{11}^{*} \right)}^{p_{e_{1}} - k}\left( t_{21}^{*} \right)^{k}\left( t_{12}^{*} %\right)^{q_{e_{2}} - l}\left( t_{22}^{*} \right)^{l}|\left( p_{e_{1}} + q_{e_{2}} - k - l %\right)_{\gamma_{1}} \right\rangle\left. \ |\left( k + l \right)_{\gamma_{2}} \right\rangle\right\rbrack
%\label{eqnc11}
%\end{eqnarray}
%\end{widetext}
\begin{eqnarray}
|p_{e_{1}}\rangle|q_{e_{2}}\rangle \Longrightarrow \frac{1}{\sqrt{p_{e_{1}}!q_{e_{1}}!}}\sum_{k = 0}^{p_{e_{1}}}\sum_{l = 0}^{q_{e_{2}}}\nonumber\\
\Bigr[\sqrt{\left(p_{e_{1}} + q_{e_{2}} - k - l \right)!\left( l + s \right)!}
\begin{pmatrix}
p_{e_{1}} \\
k \\
\end{pmatrix}
\begin{pmatrix}
q_{e_{2}} \\
l \\
\end{pmatrix}\times\nonumber\\
{\left(t_{11}^{\ast}\right)}^{p_{e_{1}} - k}\left(t_{21}^{*}\right)^{k}\left(t_{12}^{*}\right)^{q_{e_{2}} - l}\left(t_{22}^{*}\right)^{l}\times\nonumber\\
\left|\left(p_{e_{1}} + q_{e_{2}} - k - l\right)_{g_{1}}\right\rangle\left|\left(k + l\right)_{g_{2}}\right\rangle
\Bigr]
\end{eqnarray}
where $\begin{pmatrix}
p_{e_{1}} \\
k \\
\end{pmatrix}$ and $\begin{pmatrix}
q_{e_{2}} \\
l \\
\end{pmatrix}$ are binomial coefficients. It is easy to see that the
transformation preserves the number of photons in every term
$\propto |p_{e_{1}}\rangle|q_{e_{2}}\rangle$ and does not create Fock states with higher photon numbers.

Consider as an example the transformation of the single-photon state
\begin{equation}
|\Psi^{\left( e \right)}\rangle = \frac{A|1_{x}\rangle|0_{y}\rangle + B|0_{x}\rangle|1_{y}\rangle + C|0_{x}\rangle|0_{y} \rangle}{\sqrt{\left| A \right|^{2} + \left| B \right|^{2}{+ \left| C \right|}^{2}}}
\label{eqnc12}
\end{equation}
when their Cartesian basis is rotated by an angle $\phi$ around
the central ray. We assume that 
$\mathbf{e}_{1,2} \equiv \mathbf{x}_{0},\mathbf{y}_{0}$,
$\mathbf{g}_{1,2} \equiv \mathbf{x}_{0}^{\mathbf{\prime}},\mathbf{y}_{0}^{\mathbf{\prime}}$, 
and the transformation matrix is given by Eq.~(\ref{eqnc4}). As a result,
%\begin{equation}
%|\Psi^{\left( \gamma \right)}\rangle = \frac{\left(A\cos\phi + B\sin\phi\right)|1_{x^{\prime}}\rangle|0_{y^{\prime}}\rangle + \left(B\cos\phi - A\sin\phi \right)|0_{x^{\prime}}\rangle|1_{y^{\prime}}\rangle + C|0_{x^{\prime}}\rangle|0_{y^{\prime}}\rangle}{\sqrt{\left| A \right|^{2} + \left| B \right|^{2}{+ \left| C \right|}^{2}}}
%\label{eqnc13}
%\end{equation}
\begin{eqnarray}
|\Psi^{\left( g \right)}\rangle = \frac{\left(A\cos\phi + B\sin\phi\right)|1_{x^{\prime}}\rangle|0_{y^{\prime}}\rangle}{\sqrt{\left| A \right|^{2} + \left| B \right|^{2}{+ \left| C \right|}^{2}}} +
\nonumber\\ \frac{\left(B\cos\phi - A\sin\phi \right)|0_{x^{\prime}}\rangle|1_{y^{\prime}}\rangle + C|0_{x^{\prime}}\rangle|0_{y^{\prime}}\rangle}{\sqrt{\left| A \right|^{2} + \left| B \right|^{2}{+ \left| C \right|}^{2}}}
\label{eqnc13}
\end{eqnarray}

%%%%%%%%%%%%%%%%%%%%%%%%%%%%%%%%%%%%%%%%%%

\section{Interaction of a single-photon pulse with a
superposition of dark and bright states of QEs}

%%%%%%%%%%%%%%%%%%%%%%%%%%%%%%%%%%%%%%%%%%%

As we showed before, for a field with a given spatial structure one of
the QE states $|\Psi_{e_{\pm}}\rangle$ is bright, and
the second one is dark, denoted respectively as
$|\Psi_{e}\rangle$ and $|\Psi_{d}\rangle$. So far we considered the situation
when the QEs are initially either in the ground state or in the dark state. In
the first case the QEs can be excited into the bright state by a resonant photon pulse,  
whereas in the second state the pulse sees an empty cavity. Now consider
a superpositional state of the QEs,
\begin{equation}
|\Psi_{QE}\left(0\right)\rangle = G|\Psi_{g}\rangle + D|\Psi_{d}\rangle.
\label{eqnd1}
\end{equation}
Assume that a single-photon pulse incident on a cavity has a central
frequency resonant to the transition to the bright state
$|\Psi_{g}\rangle\longrightarrow|\Psi_{e}\rangle$.
The highest-energy state $|\Psi_{ee}\rangle$ will remain unoccupied because the transition
$|\Psi_{e}\rangle\longrightarrow|\Psi_{ee}\rangle$
has a different frequency; see Fig.~1. 
%We can use the
%Hamiltonian (\ref{eqn27},\ref{eqn28}) after adding the term
%\begin{equation}
%\hat{H}_{d} = W_{d}\hat{\sigma}_{d}^{\dagger}\hat{\sigma}_{d},
%\label{eqnd2}
%\end{equation}
%where ${\hat{\sigma}}_{d} = |\Psi_{g}\rangle\langle\Psi_{d}|$.
The initial quantum state of the system is given by
\begin{eqnarray}
&|\Psi\left(0\right)\rangle = \left(G|\Psi_{g}\rangle + D|\Psi_{d}\rangle\right)|0_{c}\rangle
\sum_{k}\Bigr(C_{1k}\left(0\right)|1_{1k}\rangle\prod_{k^{\prime}\neq k}\nonumber\\&\times|0_{{1k}^{\prime}}\rangle\prod_{k}{|0_{2k}\rangle} + C_{2k}\left( 0 \right)|1_{2k}\rangle\prod_{k^{\prime}\neq k}|0_{{2k}^{\prime}}\rangle\prod_{k}{|0_{1k}\rangle}\Bigr),\nonumber\\
\label{eqnd3}
\end{eqnarray}
%\begin{eqnarray}
%|\Psi\left(0\right)\rangle = \left(G|\Psi_{g}\rangle + D|\Psi_{d}\rangle\right)|0_{c}\rangle\times\\
%\sum_{k}\Bigr(C_{1k}\left(0\right)|1_{1k}\rangle\prod_{k^{\prime}\neq %k}\nonumber|0_{{1k}^{\prime}}\rangle\prod_{k}{|0_{2k}\rangle}\\ + C_{2k}\left( 0 %\right)|1_{2k}\rangle\prod_{k^{\prime}\neq k}|0_{{2k}^{\prime}}\rangle\prod_{k}{|0_{1k}\rangle}\Bigr),
%\label{eqnd3}
%\end{eqnarray}
which is normalized according to Eqs.~(\ref{eqn44}) and
\begin{equation}
\left|G\right|^{2} + \left|D \right|^{2} = 1. 
\label{eqnd4}
\end{equation}

The state vector at any moment of time can be sought as
\begin{widetext}
\begin{eqnarray}
\left|\Psi\right\rangle = &\left|\Psi_{g}\right\rangle\left|0_{c}\right\rangle\sum_{k}\left(C_{1k}\left( t \right)\left|1_{1k} \right\rangle\prod_{k^{\prime} \neq k}\left|0_{{1k}^{\prime}}\right\rangle\prod_{k}{\left|0_{2k}\right\rangle} + C_{2k}\left( t \right)\left|1_{2k}\right\rangle\prod_{k^{\prime} \neq k}\left|0_{{2k}^{\prime}}\right\rangle\prod_{k}{\left|0_{1k}\right\rangle}\right)\nonumber\\+&
\left|\Psi_{d}\right\rangle\left|0_{c}\right\rangle\sum_{k}{\left(C_{1k}^{\left( d \right)}\left( t \right)\left|1_{1k} \right\rangle\prod_{k^{\prime} \neq k}\left|0_{{1k}^{\prime}}\right\rangle\prod_{k}{\left|0_{2k}\right\rangle} + C_{2k}^{\left( d \right)}\left( t \right)\left|1_{2k}\right\rangle\prod_{k^{\prime} \neq k}\left|0_{{2k}^{\prime}} \right\rangle\prod_{k}{\left|0_{1k} \right\rangle}\right)}\nonumber\\&
+ \left(C_{c}\left( t \right)\left|\Psi_{g} \right\rangle + C_{c}^{\left( d \right)}\left( t \right)\left|\Psi_{d} \right\rangle \right)\left|1_{c} \right\rangle\prod_{k}{\left|0_{1k}\right\rangle}\prod_{k}{\left|0_{2k}\right\rangle} + \nonumber
C_{e}\left( t \right)\left|\Psi_{e} \right\rangle\left|0_{c} \right\rangle\prod_{k}{\left|0_{1k} \right\rangle\prod_{k}{\left|0_{2k} \right\rangle}}\\& + \left( C_{0}\left( t \right)\left|\Psi_{g} \right\rangle + C_{0}^{\left( d \right)}\left( t \right)\left|\Psi_{d} \right\rangle \right)\left|0_{c} \right\rangle\prod_{k}{\left|0_{1k} \right\rangle}\prod_{k}{\left|0_{2k} \right\rangle}.
\label{eqnd5}
\end{eqnarray}
\end{widetext}
We need to solve Eqs.~(\ref{eqn30})-(\ref{eqn34}) for the complex probability
amplitudes after adding to them a set of equations for $C_{1k}^{\left(d\right)}$,
$C_{2k}^{\left(d\right)}$, $C_{c}^{\left(d\right)}$,
and $C_{0}^{\left(d\right)}$. These two sets are not coupled to
each other if the pulse duration is shorter that the decay time of the
dark state. At the same time, the mere presence of the dark state
$\left|\Psi_{d} \right\rangle$, which can be occupied at the
initial moment of time with probability $\left| D \right|^{2}$,
changes the normalization condition in Sec.~III to
\begin{eqnarray}
\overline{C_{0}\left( t \right) - C_{0}\left( 0 \right)} = 0,\\
\overline{\left| C_{0} \right|^{2} + \sum_{k}\left( \left| C_{1k} \right|^{2} + \left| C_{2k} \right|^{2} \right) + \left| C_{c} \right|^{2} + \left| C_{e} \right|^{2}} = \nonumber\\1 - \left|D\right|^{2} = \left|G\right|^{2}.
\label{eqnd6}
\end{eqnarray}
Equations of motion for $C_{1k}^{\left( d \right)}$,
$C_{2k}^{\left( d \right)}, C_{c}^{\left( d \right)}$,
and $C_{0}^{\left( d \right)}$ are similar to those for $C_{1k}$,
$C_{2k}$, $C_{c}$, $C_{0}$ in the limiting case
$\left| \Omega_{c} \right| = 0$; it has to be taken into account as
well that the dark state is excited with probability $\left| D \right|^{2}$:
\begin{equation}
\frac{dC_{1k}^{\left(d \right)}}{dt} + i\left(\omega_{k} + \frac{W_{d}}{\hbar} \right)C_{1k}^{\left( d \right)} - i\mathfrak{L}^{*}C_{c}^{\left( d \right)} = 0
\label{eqnd7}
\end{equation}
\begin{equation}
\frac{dC_{c}^{\left( d \right)}}{dt} + \left( i\omega_{c} + \frac{W_{d}}{\hbar} + \frac{\mu_{c}}{2} \right)C_{c}^{\left( d \right)} - i\sum_{k}{\mathfrak{L}C_{1k}} = 0
\label{eqnd8}
\end{equation}
\begin{equation}
\frac{dC_{2k}^{\left( d \right)}}{dt} + i\left( \omega_{k} + \frac{W_{d}}{\hbar} \right)C_{2k}^{\left( d \right)} = 0,
\label{eqnd9}
\end{equation}
\begin{eqnarray}
\overline{C_{0}^{\left( d \right)}\left( t \right) - C_{0}^{\left( d \right)}\left( 0 \right)} = 0,\nonumber\\
\overline{\left| C_{0}^{\left( d \right)} \right|^{2} + \sum_{k}^{}\left( \left| C_{1k}^{\left( d \right)} \right|^{2} + \left| C_{2k}^{\left( d \right)} \right|^{2} \right) + \left| C_{c}^{\left( d \right)} \right|^{2}} = \left| D \right|^{2}.\nonumber\\
\label{eqnd10}
\end{eqnarray}
Using the principle of superposition and the expressions obtained in
previous sections, we can write down the solution for the state vector
similarly to Eq.~(\ref{eqn68}). Dropping a small
contribution due to fluctuations caused by pure dephasing of the bright
state, we obtain
\begin{widetext}
\begin{eqnarray}
\left|\Psi\left( t > T_{1} \right) \right\rangle =&
G\left|\Psi_{g}\right\rangle\left|0_{c}\right\rangle\sum_{k}{e^{-i\omega_{k}t}\Bigr({R_{1}}C_{1k}\left(0 \right)\left|1_{1k}\right\rangle\prod_{k^{\prime} \neq k}\left|0_{{1k}^{\prime}}\right\rangle\prod_{k}{\left|0_{2k} \right\rangle} + C_{2k}\left( 0 \right)\left|1_{2k} \right\rangle\prod_{k^{\prime} \neq k}\left|0_{{2k}^{\prime}} \right\rangle\prod_{k}{\left|0_{1k} \right\rangle}\Bigr)}\nonumber +\\De^{-i\frac{W_{d}}{\hbar}t}&
\left|\Psi_{d}\right\rangle\left|0_{c}\right\rangle\sum_{k}{e^{- i\omega_{k}t}\Bigr({R_{1}}_{\left| \Omega_{c} \right| = 0}C_{1k}\left( 0 \right)|1_{1k}\rangle\prod_{k^{\prime} \neq k}\left|0_{{1k}^{\prime}}\right\rangle\prod_{k}{\left|0_{2k} \right\rangle} + C_{2k}\left( 0 \right)\left|1_{2k} \right\rangle\prod_{k^{\prime} \neq k}\left|0_{{2k}^{'}}\right\rangle\prod_{k}{\left|0_{1k} \right\rangle} \Bigr)}\nonumber\\& +
 \left(C_{0}\left|\Psi_{g} \right\rangle + C_{0}^{\left( d \right)}{e^{- i\frac{W_{d}}{\hbar}t}|\Psi}_{d}\rangle \right)\left|0_{c} \right\rangle\prod_{k}{\left|0_{1k} \right\rangle}\prod_{k}{\left|0_{2k}\right\rangle},
\label{eqnd11}
\end{eqnarray}
\end{widetext}
where the factor 
${R_{1}}_{\left| \Omega_{c} \right| = 0}$ is $R_1$ from Eq.~(\ref{eqn69}) evaluated at $\left| \Omega_{c} \right| = 0$;  
\begin{equation}
\overline{C_{0}} = 0,\
\overline{\left| C_{0} \right|^{2}} = \left|G \right|^{2}\left\lbrack 1 - \sum_{k}\left(\left|{R_{1}}C_{1k}\left( 0 \right) \right|^{2} + \left| C_{2k}\left( 0 \right) \right|^{2} \right) \right\rbrack,
\label{eqnd12}
\end{equation}
\begin{eqnarray}
\overline{C_{d}} = 0,\
\overline{\left|C_{0}^{\left( d \right)} \right|^{2}} = \left| D \right|^{2}\times\nonumber\\\left\lbrack 1 - \sum_{k}\left(\left|{R_{1}}_{\left|\Omega_{c}\right| = 0}C_{1k}\left( 0 \right) \right|^{2} + \left| C_{2k}\left( 0 \right) \right|^{2} \right) \right\rbrack.
\label{eqnd13}
\end{eqnarray}
In particular, for $\Delta_{0} = 0$,
$\left|\Omega_{c}\right| \gg \kappa_{\Sigma},\ \left| p_{e} \right|$
and $\frac{\mu_{c}}{2} \ll \kappa$ we get
\begin{equation}
{R_{1}} \approx 1, \; 
{R_{1}}_{\left| \Omega_{c} \right| = 0} \approx - 1; 
\label{eqnd14}
\end{equation}
\begin{equation}
\overline{\left| C_{0} \right|^{2}} \approx \overline{\left| C_{0}^{\left( d \right)} \right|^{2}} \approx 0.
\label{eqnd15}
\end{equation}
Considering a single-mode incident field for simplicity as in Sec.~IV
and omitting unimportant phase factors, we arrive at
%\begin{eqnarray}
%\left|\Psi\left(t > T_{1}\right)\right\rangle = &&\left|0_{c}\right\rangle\left\lbrack G\left|\Psi_{g} &\right\rangle\left(C_{1}\left|1_{1} \right\rangle\left|0_{2} \right\rangle + C_{2} \left|0_{1} &\right\rangle\left|1_{2} \right\rangle \right)+\nonumber\\ &&D\left|\Psi_{d}\right\rangle\left(-&C_{1}\left|1_{1}\right\rangle\left|0_{2} \right\rangle + C_{2}\left|0_{1}\right\rangle\left|1_{2} &\right\rangle \right) \right\rbrack
%\label{eqnd16}
%\end{eqnarray}
\begin{eqnarray}
\left|\Psi\left(t > T_{1}\right)\right\rangle &=& \left|0_{c}\right\rangle \Big[ G\left|\Psi_{g}\right\rangle\left(C_{1}\left|1_{1}\right\rangle\left|0_{2}\right\rangle + C_{2}\left|0_{1}\right\rangle\left|1_{2}\right\rangle\right) \nonumber\\
&& + D\left|\Psi_{d}\right\rangle\left(-C_{1}\left|1_{1}\right\rangle\left|0_{2}\right\rangle + C_{2}\left|0_{1}\right\rangle\left|1_{2}\right\rangle\right) \Big]\nonumber\\
\label{eqnd16}
\end{eqnarray}
where the indices 1 and 2 in Eqs. (\ref{eqnd16}) and (\ref{eqnd17}) correspond to the
Cartesian axes which define the orientation of the illuminated cavity
surface (see Sec.~IV) and real constants $C_{1,2}$ correspond to the
initial state
\begin{eqnarray}
\left|\Psi\left(t = 0\right)\right\rangle &= &\left|0_{c}\right\rangle\left(G\left|\Psi_{g}\right\rangle + D\left|\Psi_{d}\right\rangle\right)\times\nonumber\\&&\left(C_{1}\left|1_{1}\right\rangle\left|0_{2}\right\rangle + C_{2} \left|0_{1}\right\rangle\left|1_{2}\right\rangle\right).
\label{eqnd17}
\end{eqnarray}

As we saw in Sec.~IV, for real $C_{1,2}$ the transformation
\begin{equation}
C_{1}\left|1_{1} \right\rangle\left|0_{2}\right\rangle + C_{2}\left|0_{1}\right\rangle\left|1_{2}\right\rangle \Longrightarrow - C_{1}\left|1_{1}\right\rangle\left|0_{2}\right\rangle + C_{2}\left|0_{1} \right\rangle\left |1_{2} \right\rangle
\label{eqnd18}
\end{equation}
corresponds to the rotation of the plane of linear polarization of an
incident photon by an angle $2\phi$, where $\phi$ is an angle
between the polarization plane of the incident photon and the axis of
the cavity, as in Fig.~2(a). 

At the same time, there is an important difference between the situation
considered in Sec.~IV and the present case. When the incident photon
interacts with QEs initially in the ground or dark state, i.e., when
$D = 0$ or $G = 0$ in Eqs.~(\ref{eqnd16}) and (\ref{eqnd17}), the polarization plane is
preserved upon reflection or is rotated, but the reflected photon remains
linearly polarized. In other words, with proper orientation of the
polarization filter in front of the detector there is always a plane for
which the probability of the photon detection is zero. However, if the
QEs were initially in the superposition state, the reflected photon
remains linearly polarized only if $C_{2} = 0$ or $C_{1} = 0$, when
the state vector (\ref{eqnd16}) has the form
\begin{equation}
\left|\Psi\left(t > T_{1} \right)\right\rangle = \left|0_{c}\right\rangle\left(G\left|\Psi_{g} \right\rangle - D\left|\Psi_{d} \right\rangle\right)C_{1}\left|1_{1}\right\rangle\left|0_{2} \right\rangle\nonumber
\end{equation}
or
\begin{equation}
\left|\Psi\left( t > T_{1} \right) \right\rangle = \left|0_{c} \right\rangle\left(G\left|\Psi_{g}\right\rangle + D\left|\Psi_{d} \right\rangle\right)C_{2}\left|0_{1}\right\rangle\left|1_{2} \right\rangle, 
\label{eqnd19}
\end{equation}
respectively. In these two cases the polarization plane of the incident
photon coincides with one of the symmetry axes of the cavity. In the
general case, the reflected photon becomes only partially polarized as a
result of entanglement of quantum states of the photon and QEs. For
example, if $\phi = \frac{\pi}{4}$, then in the coordinate system in
which the photon state has the form 
$\left|1_{x}\right\rangle\left|0_{y} \right\rangle$, the
state vector (\ref{eqnd16}) becomes
\begin{equation}
\left|\Psi\left(t > T_{1}\right)\right\rangle = \left|0_{c} \right\rangle\left( G\left|\Psi_{g} \right\rangle\left|1_{x} \right\rangle\left|0_{y}\right\rangle + D\left|\Psi_{d} \right\rangle\left|0_{x} \right\rangle\left|1_{y} \right\rangle \right). 
\label{eqnd20}
\end{equation}
In this state the probability of detecting a photon with polarization
making an angle $\vartheta$ with the \emph{x} axis is equal to
$\frac{2}{\pi}\left\lbrack\left|G\right|^{2}\cos^{2}\left(\vartheta\right) + \left|D\right|^{2}\sin^{2}\left(\vartheta\right) \right\rbrack$,
where
$\vartheta \in \left\lbrack \frac{\pi}{2}, - \frac{\pi}{2}\  \right\rbrack$.
In particular, for $\left|G\right|^{2} = \left| D \right|^{2}$ all
orientations of linear polarization are equally probable. The classical
analog of this state is the superposition of two incoherent linearly
polarized fields.

%%%%%%%%%%%%%%%%%%%%%%%%%%%%%%%%%%%%

\end{document}